\documentclass[aps,prd,twocolumn,floatfix,preprintnumbers,nofootinbib]{revtex4-1}

\usepackage[margin=0.7in]{geometry}
\usepackage[utf8]{inputenc}

\usepackage{physics}
\usepackage{amsmath,amssymb,amsthm,amsfonts}
\usepackage[final]{graphicx}
\usepackage{subcaption}
\usepackage{lipsum}
\usepackage{mathtools}
\usepackage{enumitem,tabulary}
\usepackage{indentfirst}
\usepackage[english]{babel}
\usepackage{textcomp}
\usepackage{multirow}
\usepackage{tikz}
\usepackage{tipa}
\usepackage{CJKutf8}
\usepackage{feynmf}
\usepackage{slashed}
\usepackage{braket}
\usepackage[toc,page]{appendix}
\usepackage{url}
\usepackage{natbib}
\usepackage{graphicx}

\usepackage{mathrsfs}  
\usepackage{cancel}
\usepackage[normalem]{ulem}
\usepackage{array}
\usepackage{booktabs}
\usepackage{verbatim}
\usepackage{ragged2e}

\usetikzlibrary{positioning,decorations.pathmorphing,decorations.markings,arrows}
\usepackage{latexsym,amsfonts,color,amsthm}
\usepackage[compat=1.1.0]{tikz-feynman}
\RequirePackage[colorlinks=true
,urlcolor=blue
,anchorcolor=blue
,citecolor=blue
,filecolor=blue
,linkcolor=blue
,menucolor=blue
,linktocpage=true
,pdfproducer=medialab
,pdfa=true
]{hyperref}

\newcommand\fy[1]{\slashed{#1}}

\DeclareMathOperator{\arccosh}{arccosh}

\begin{document}
\begin{flushright}
MI-HET-809
\end{flushright}

\title{Longer-Lived Mediators from Charged Mesons and Photons at Neutrino Experiments}

\author{Bhaskar Dutta}
\email{dutta@tamu.edu}
\affiliation{Department of Physics and Astronomy,
Texas A\&M University, College Station, TX-77845 USA}

\author{Aparajitha Karthikeyan}
\email{aparajitha\_96@tamu.edu}
\affiliation{Department of Physics and Astronomy,
Texas A\&M University, College Station, TX-77845 USA}

\author{Doojin Kim}
\email{doojin.kim@tamu.edu}
\affiliation{Department of Physics and Astronomy,
Texas A\&M University, College Station, TX-77845 USA}

\begin{abstract}
    Since many of the dark-sector particles interact with Standard Model (SM) particles in multiple ways, they can appear in experimental facilities where SM particles appear in abundance. In this study, we explore a particular class of longer-lived mediators that are dominantly produced from photons and charged mesons that arise in proton-beam fixed-target-type neutrino experiments.
    This class of mediators encompasses light scalars that appear in theories like extended Higgs sectors, muon(electro)philic scalars, etc.
    We evaluate the sensitivities of these mediators at beam-based neutrino experiments such as the finished ArgoNeuT, ongoing MicroBooNE, SBND, ICARUS, and the upcoming DUNE experiment. We find that muonphilic scalars are more enhanced while produced from three-body decay of charged mesons.
    The above-mentioned experiments can probe unexplored regions of parameter space that can explain the current discrepancy in the anomalous magnetic moment of muons. We further find that Compton-like scattering of photons is the largest source of electrophilic scalars. By utilizing this, the DUNE Near Detector can explore new regions in the sensitivity space of electrophilic scalars. We also show that Bethe-Heitler scattering processes can be used to probe flavor-specific lepton final states even for the mediator masses below twice the lepton mass.

\end{abstract}

\maketitle

\tableofcontents

\section{Introduction}

There are compelling reasons for the existence of a particle sector (often called dark sector or hidden sector) beyond the Standard Model (SM) of particle physics, for example, dark matter, non-zero neutrino masses, mass-flavor hierarchy puzzle, etc. 
An attractive scenario consists of a new particle sector that is very weakly or feebly connected to the SM sector via portal particles that are often called mediators~\cite{Batell:2022xau}.
A myriad of models with mediators has been built along this line to address these issues and similarly, many experiments are being developed towards unraveling these mysteries. 
These efforts are also motivated 
by anomalies in the experimental results such as the LSND excess \cite{LSND:2001aii}, the MiniBooNE anomaly \cite{MiniBooNE:2008yuf, MiniBooNE:2018esg, MiniBooNE:2020pnu}, and the discrepancy in the anomalous magnetic moments of the muon~\cite{Muon:g-21,Muon:g-22} and electron~\cite{g-2e}. 
A subset of these experiments are fixed-target-type experiments involving high-intensity protons on target (POT) and they are widely adopted at neutrino facilities with beam energy being $O(1)$ to a few hundred GeV.
While neutrino facilities serve as neutrino factories, copious amounts of charged mesons, (secondary) photons, electrons, positrons, and neutral mesons are also produced. Therefore, given the beam energy and intensity of neutrino facilities, they can test MeV-to-sub-GeV-scale new physics particles interacting with those SM particles that can be found at these facilities. 

While the landscape of dark-sector models is vast, we will focus on a particular class of models that constitutes scalar mediators that couple either to all SM matter particles or a subset of flavors. 
We find that the flux of the above mediators produced from charged mesons inside a proton target can be quite enhanced. Scalars could abundantly appear through the three-body decay of charged mesons such as charged pions and kaons; their corresponding two-body decay is helicity-suppressed but adding another particle in the final state would evade the suppression and enhance the branching fraction of three-body decay modes~\cite{rislow:3body}. We also notice an enhancement when they are sourced from photons. 
For example, scalars can be copiously produced as a consequence of Primakoff scattering~\cite{primakoff,Dent:2019ueq,Brdar:2020dpr} of photons as they interact with atoms present in the target. 
This process is coherently enhanced by a factor of $Z^2$ from the nuclear form factor~\cite{tsai:prim}. 
Similar to scalars, flavor-specific massive vector mediators can be produced from the above sources as well. One such example we considered in this paper is a muonphilic gauge boson that appears in the context of a $U(1)_{T3R}$ model~\cite{Dutta:2019fxn, Dutta:2020jsy, Dutta:2020enk, Dutta:2022qvn}. We find that these can also appear in good abundance from charged mesons, neutral mesons, photons, {electrons, and bremsstrahlung}. 

We investigate the detection prospects of the signals induced by the aforementioned mediators at experiments along the NuMI~\cite{Adamson:2015dkw} and BNB~\cite{Machado:2019oxb} beamlines at Fermi National Laboratory: ArgoNeuT~\cite{Anderson:2012vc}, ICARUS~\cite{ICARUS:2004wqc}, MicroBooNE~\cite{MicroBooNE:2016pwy}, and SBND~\cite{MicroBooNE:2015bmn}. 
We also probe them at the upcoming DUNE Near Detector~\cite{DUNE:2021tad} (DUNE ND), which is placed along the LBNF beam line~\cite{DUNE:2016hlj}. These detectors feature different baselines and angular distances with respect to their respective beam axis. The magnetic horns -- which are designed to focus or deflect charged particles and, in turn, their corresponding neutrino decay products -- affect the mediator production via charged mesons. As a result of the magnetic horn effect along with the position of detectors, we expect to take advantage of the multiple experiments and probe regions of parameter space in a complementary manner.

Once a mediator is produced inside the proton target of these beam facilities, it should be safely delivered to a detector of interest. 
Due to the feebly-interacting nature of the above-mentioned mediators, they would live longer rather than decay immediately. 
Once they survive, some fraction of mediators can leave detectable signatures within the detector fiducial volume. These signatures include electron-positron pairs, muon-antimuon pairs, photon pairs, electron-photon pairs, and single photons from scattering and decay processes. We again expect to benefit from different baselines and detector sizes in the search for these long-lived mediators as they provide complementarity. 

We emphasize that while one can look for electron-positron pairs and muon-antimuon pairs from decay processes, the same final states can arise through the splitting process, also known as the Bethe-Heitler scattering process~\cite{Bethe:1934za}. 
Owing to its energy-dependent nature, these final states can also be produced from a mediator with a keV-range mass, which is not kinematically allowed for decay processes. 
For example, a 10 keV (muonphilic) scalar with a total energy greater than 210 MeV can split into a muon-antimuon pair through the Bethe-Heitler scattering process. 
From the above example, we see that the appearance of lepton-antilepton final states for all possible masses helps us to probe flavor-specific models directly. 
With all the above-mentioned detection channels, we investigate the parameter space of the Higgs Portal Scalar and $(g-2)$-motivated parameter space, utilizing muon/electrophilic scalar mediators which can be efficiently probed by experiments operating at the sub-GeV scale.

In Sec.~\ref{SectionModels}, we discuss essential features of the example models that we explore. Section~\ref{SectionExperiments} is reserved for a brief overview of the benchmark short baseline experiments for which we study sensitivity reaches. 
We then explain how the aforementioned mediators are produced at generic proton-on-target experiments in Sec.~\ref{SectionProduction} and elaborate on the signals that the mediators produce at the detectors in Sec.~\ref{Signals}. In Sec.\ref{SectionSensitivities}, we explain our analysis methodology and report our main results including sensitivity plots. In Sec.\ref{sec:vectors}, we explore the above study for vector mediator scenarios using the $U(1)_{T3R}$ model as an example. We finally summarize and conclude our study in Sec.~\ref{sec:conclusions}.

\section{Models} \label{SectionModels}
We apply the above production and detection mechanisms of scalars in the context of three benchmark spin-0 mediator models.\footnote{We emphasize that our study here can be straightforwardly applied to mediators with different Lorentz structures as well.} Although we focus on scalars in the paper, one can also look at spin-1 gauge boson mediator models. We will briefly explore these aspects in Sec.~\ref{sec:vectors}.  

\subsection{Higgs Portal Scalars (HPS)}
This model contains a singlet dark scalar $S$ with mass $m_S$ that interacts with the SM $SU(2)$ scalar doublet $H$ via a portal interaction \cite{PattHPS:2006}:
\begin{equation}
    \mathcal{L}_{S} \supset (AS + BS^2)H^{\dagger}H.
    \label{eq:HPS}
\end{equation} 
Where $A$ and $B$ are free parameters. Under $SU(2)_L$ symmetry breaking, the neutral Higgs decomposes into a sum $v + h$. 
Therefore, the interaction in Eq.~\eqref{eq:HPS} induces a mass mixing between the dark scalar and the SM Higgs in two ways: ($a$) if $A\neq 0$, then the mass mixing is naturally induced regardless of whether the dark scalar acquires a zero or a non-zero dark scalar vacuum expectation value (vev), and ($b$) if $A=0$, the dark scalar can acquire a non-zero vev by an appropriate choice of potential and thus induce mass mixing. After diagonalizing the mass-like terms, we see that the scalar $S$ mixes with the SM Higgs $h$ via a small mixing angle, that is,
\begin{equation}
    h \rightarrow h + \theta S.
\end{equation}
Therefore, the dark scalar can interact with the SM particles that acquire mass via the Higgs scalar:
\begin{equation}
    \mathcal{L_{S}} \supset \frac{1}{2}m_S^2 S^2 + \theta S \big( \sum_{f} \frac{m_f}{v} \bar{f}f + \frac{2m_W^2}{v}W^+_{\mu}W_{-}^{\mu} + \frac{m_Z^2}{v}Z_{\mu}Z^{\mu} \big),
\end{equation}
where $f$ runs over all quark and charged lepton flavors.

This model is of particular interest in various contexts. Examples include scalar-to-pion decays~\cite{Gorbunov:2023lga},  MicroBooNE searches for the HPS to explain the KOTO excess~\cite{MicroBooNE:2021usw,MicroBooNE:2022ctm}, and a search for the HPS-induced signatures at ICARUS~\cite{BatellHPS:2019}. 
In addition to the above-shown interactions, HPS can also couple to two photons via a fermion loop and thus widen the phenomenology \cite{Cosme:2018nly}.

\subsection{Muonphilic scalars}
These scalars (henceforth denoted by $\phi_{\mu}$) can appear in effective field theories containing singlet scalars that have minimal flavor violation~\cite{BatellScalar:2017} or in other extensions to the SM~\cite{Harris:2022vnx} with additional doublets/singlets, which contain Yukawa couplings unique to each flavor~\cite{DuttaScalars:2020}. 
The Lagrangian has the following form. 
\begin{equation}
    \mathcal{L}_{\phi,\mu} \supset y_{22} \bar{\mu}\mu \phi_{\mu}.
\end{equation}
Muonphilic scalars contribute to the anomalous magnetic moment of the muon ($a_{\mu}$) of the muon at the one-loop level. Therefore, their phenomenology is useful to explain the $4.2\sigma$ discrepancy between the experimental and theoretical values of $a_{\mu}$~\cite{Muon:g-21,Muon:g-22}:
\begin{equation}\label{g-2mueq}
    \Delta a_{\mu} = a_{\mu}^{\rm exp} - a_{\mu}^{\rm th} = (2.51 \pm .59) \times 10^{-9}.
\end{equation} 

Similar phenomenology has been explored via muon-coupled axion models which bring in constraints from SN1987a data~\cite{Croon:2020lrf, Caputo:2021rux}. 
These scalars can also couple to two photons via a muon loop and the effect of this has been studied in the context of axions~\cite{Dobrich:2015jyk,Dolan:2017osp}. 
The lack of photon events at the E137 SLAC experiment imposes stringent constraints on this model as well~\cite{E137}.

\subsection{Electrophilic scalars}
On a similar line of thought, there are models with scalars (henceforth denoted by $\phi_e$) that solely couple to electrons. One such example is an effective field theory where all heavy fermions and bosons are integrated out such that we end up with scalars that exclusively couple to electrons via a Yukawa coupling. Such a model has bounds from stellar cooling \cite{Hardy:2016kme, Bottaro:2023gep}, SN1987a \cite{SN1987a:axion}, NA64 \cite{NA64e}, Orsay \cite{Orsay}, E141 \cite{E141} and E774 \cite{E774} that look at electron-positron as well as electron-photon final states. The relevant Lagrangian is given by
\begin{equation}
    \mathcal{L}_{\phi,e} \supset y_{11} \bar{e}e \phi_e.
\end{equation}
Like $\phi_{\mu}$ searches, we can study electrophilic scalars to explain the discrepancy in the electron anomalous magnetic moment $a_{e}$, which, based on a recent measurement with $^{87}\text{Rb}$~\cite{g-2e}, is
\begin{equation}\label{g-2eeq}
    \Delta a_{e} = a_{e}^{\rm exp} - a_{e}^{\rm th} = (4.8 \pm 3.0) \times 10^{-13}\,.
\end{equation}

\section{Benchmark Experiments}\label{SectionExperiments}

We explore the sensitivity of these models in several neutrino experiments as mentioned earlier. We tabulate key specifications of the experiments in Table.~\ref{ExpTables}. 
For the ease of the simulation, we simplify the detector geometry to cylinders. We hence specify the dimensions for the full cube-shaped detector as well as the cylinder-shaped one that is used in our simulations. Nevertheless, we expect that our main conclusions are nearly unaffected by these changes as the modified volumes fit in the original ones.
The aforementioned mediators that reach ArgoNeuT, ICARUS, and MicroBooNE are sourced from the 120~GeV NuMI beam, those at SBND are from the 8~GeV BNB beam, and finally, those at DUNE ND are produced by the 120~GeV LBNF beam. MicroBooNE and ICARUS receive a considerable number of mediators from the BNB beam on account of being placed on its axis. In this study, however, we will consider the contributions from the NuMI beam only.

\begin{table*}[t]
\resizebox{2\columnwidth}{!}{
    \begin{tabular}{| m{3cm} || m{2cm} | m{1.5cm} | m{2cm} | m{3.5cm} | m{3cm} | m{2cm} |}
        \hline
        Detectors & Beam, Energy & Distance  & Angle off-axis & Dimensions ($l \times w \times d$) & Dimensions ($r \times d$) & POTs\\
         & [GeV] & [m] & [degrees] & [m $\times$ m $\times$ m] & [m $\times$ m] for sim. & \\
        \hline 
        SBND~\cite{MicroBooNE:2015bmn} & BNB, 8 & 110 & 0.3 & $4\times4\times5$ & $2.25 \times 5$ & $6.6 \times 10^{20}$ \\
        DUNE ND~\cite{DUNE:2021tad} & LBNF, 120 & 574 & 0 & $3\times5\times 4$ & $2.18 \times 4$ & $7 \times 10^{21}$ \\
        ArgoNeuT~\cite{Anderson:2012vc} & NuMI, 120 & 1040 & 0 & $0.4 \times 0.48 \times 0.9$ & $0.25 \times 0.9$ & $1.35 \times 10^{20}$ \\
        ICARUS~\cite{ICARUS:2004wqc} & NuMI, 120 & 803 & 5.56 &  $2 \times (2.63 \times 2.86 \times 17$) & $3.10 \times 17$ & $6.6 \times 10^{20}$ \\
        MicroBooNE~\cite{MicroBooNE:2016pwy} & NuMI, 120 & 685 & 8 & $2.26 \times 2.03 \times 10.4$ & $1.21 \times 10.4$ & $6 \times 10^{20}$ \\
        \hline 
    \end{tabular}
    }
    \captionsetup{justification=Justified}
    \caption{List of experiments at the NuMI, BNB, and LBNF baselines and key specifications of the detectors. The dimensions in the second last column are the adopted ones for simplified simulation purposes. The numbers quoted in the last column are the POT that we use in our study.}
    \label{ExpTables}
\end{table*}

\begin{figure}[h]
     \begin{subfigure}{.48\textwidth}
        \centering
        \includegraphics[width=\textwidth,clip]{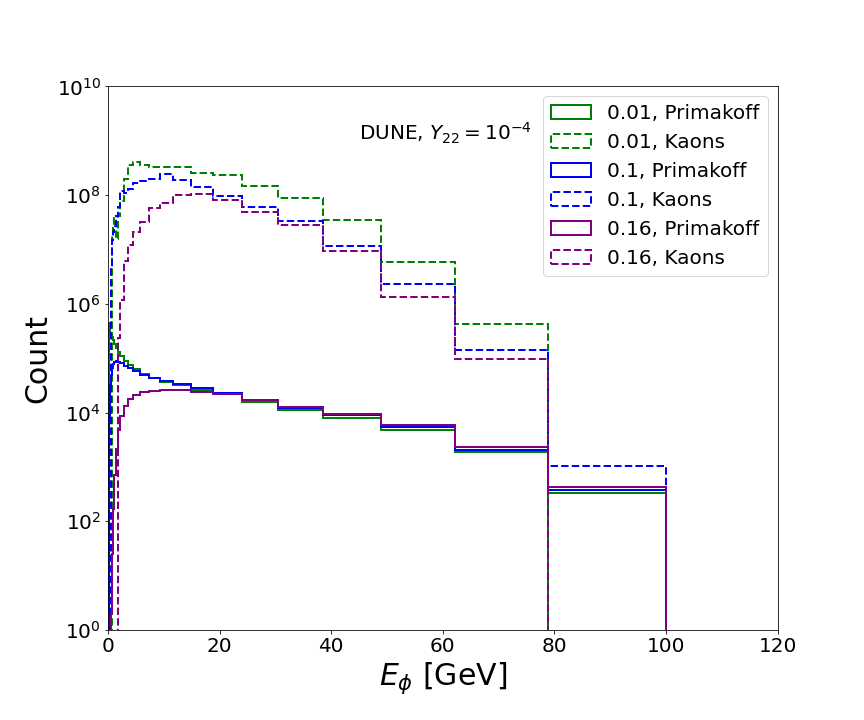}
         \captionsetup{justification=justified, singlelinecheck=false}
         \caption{Flux of muonphilic scalars that reach DUNE ND}\label{fig:energyflux1}
     \end{subfigure}\\
     \begin{subfigure}{.48\textwidth}
        \centering
         \includegraphics[width=\textwidth,clip]{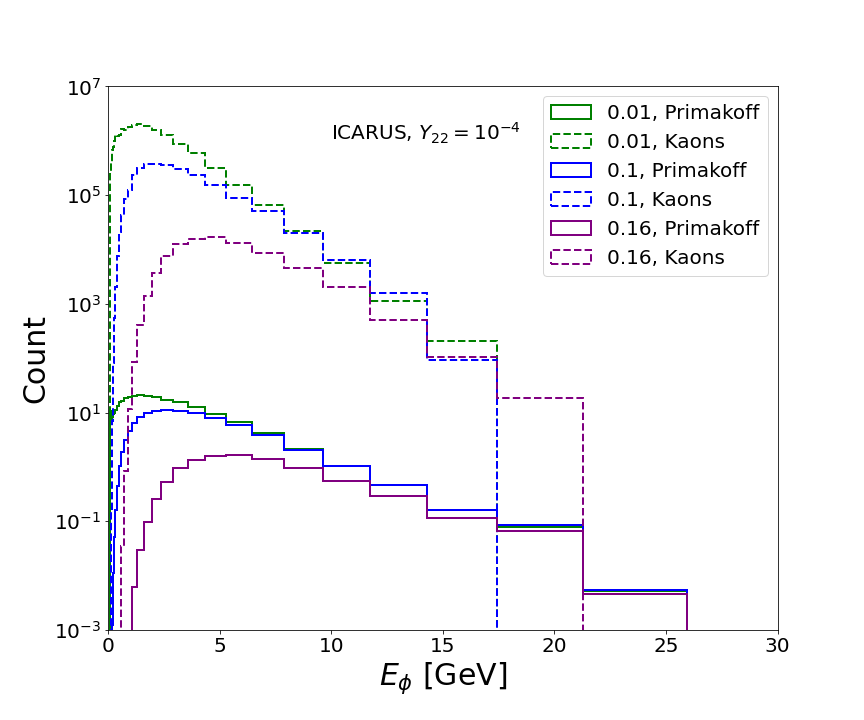}
         \captionsetup{justification=justified, singlelinecheck=false}
         \caption{Flux of muonphilic scalars that reach ICARUS}\label{fig:energyflux2}
     \end{subfigure}
     \captionsetup{justification=Justified, singlelinecheck=false}
     \caption{Energy spectra of muonphilic scalars at (a) DUNE ND and (b) ICARUS. 
     Three scalar mass values are shown: $m_{\phi} = 0.01$~\text{GeV}, 0.1~GeV, and 0.16~GeV with Yukawa coupling $Y_{22} = 10^{-4}$. }
     \label{fig:energyflux}
 \end{figure}

The magnetic horn system present near the target plays an integral role. They operate in either neutrino mode (focusing positive mesons) or antineutrino mode (focusing negative mesons).
We remark that DUNE ND, ArgoNeuT, and SBND are located on the beamline,\footnote{The BNB beam axis gets through SBND and the detector center is off the beam axis by 0.3 degrees.} whereas ICARUS and MicroBooNE are placed off-axis. Since the magnetic horns focus charged particles along the beam axis, most of the high-energy mediators (originating from high-energy charged mesons) are boosted along the beam direction, whereas softer mediators are less focused and diverge away from the axis. Hence, on-axis detectors are more sensitive to high-energy mediators as compared to those that are off-axis. 
Similarly, (secondary) high-energy photons are directed more in the forward direction and softer photons are directed more away from the beam axis. Therefore, high-energy mediators are likely to travel along the beam axis. These expectations are depicted in Fig.~\ref{fig:energyflux} which contains the energy spectra of muonphilic scalars at DUNE ND (see Fig.~\ref{fig:energyflux1}), which is one of the on-axis detectors, and ICARUS (see Fig.~\ref{fig:energyflux2}), which is off-axis. 
We see that the energies of the scalars that reach ICARUS are much lower than those at DUNE ND.
 
\section{Production of mediators} \label{SectionProduction}

In this section, we explain how long-lived scalars can be produced from charged mesons, photons, electrons, and positrons.

\subsection{Charged meson decays}

\begin{figure}[t]
    \centering
    \subfloat[]
    {
    \begin{tikzpicture}
        \begin{feynman}
            \vertex (a) at (-1.3,0) {$K^{+}$ };
            \vertex (b) at (0,0);
            \vertex (c) at (0.5,0.5);
            \vertex (d) at (1.1,1.1) {$\ell^{+}$};
            \vertex (e) at (1,-1) {$\nu_{\ell}$};
            \vertex (f) at (1.3,-0.3) {$\phi/S$};
            \diagram*{
            (a) -- [scalar, very thick](b) -- [fermion, very thick](d),
            (b) -- [fermion, very thick](e),
            (c) -- [scalar, very thick] (f),
            };
        \end{feynman}
    \end{tikzpicture}
    \label{fig:leptonscalar}
    }
    \vspace{0.3cm}
    \subfloat[]
    {
    \begin{tikzpicture}
        \begin{feynman}
            \vertex (a) at (-1.2,0.7){$u$};
            \vertex (b) at (0,0);
            \vertex (c) at (-0.3,0.7);
            \vertex (d) at (-1.2,-0.7) {$\bar{s}$};
            \vertex (e) at (-0.3,-0.7);
            \vertex (f) at (0.65,1) {$S$};
            \vertex (g) at (0.65,0);
            \vertex (v) at (1.3,0);
            \vertex (l) at (2.3,0.9){$\ell^+$};
            \vertex (n) at (2.3,-0.9){$\nu_{\ell}$};
            \node at (0.7, -0.4) {$W^+$};
            \diagram*{
            (a) -- [fermion, very thick](c) -- [fermion, very thick, quarter left](b) -- [fermion, very thick, quarter left](e) -- [fermion, very thick] (d),
            (b) -- [boson, very thick](v),
            (g) -- [scalar, very thick] (f),
            (l) --[fermion,very thick](v) -- [fermion, very thick] (n)
            };
        \end{feynman}
    \end{tikzpicture}
    \label{fig:wbosonscalar}
    }
    \hspace{1cm}
    \subfloat[]
    {
    \begin{tikzpicture}
        \begin{feynman}
            \vertex (a) at (-2,0.5){$\bar{s}$};
            \vertex (b) at (-0.8,0.5);
            \vertex (c) at (0.8,0.5);
            \vertex (d) at (2,0.5){$\bar{u}$};
            \vertex (e) at (-2,-0.5){$d$};
            \vertex (f) at (2,-0.5) {$d$};
            \vertex (g) at (0.5, 1.5) {$S$};
            \vertex (h) at (0, 0.5) ;
            \node at (0, 0.25) {$\bar{t}$};
            \node at (-2.5, 0) {$K^+$};
            \node at (2.5, 0) {$\pi^+$};
            \diagram*{
            (a) -- [fermion, thick](b) -- [fermion, thick, label = {[yshift =1cm]$\bar{t}$}](c) -- [fermion,very thick](d),
            (b) -- [boson, half right, looseness = 1.5](c),
            (e) --[fermion, thick](f),
            (h) -- [scalar, thick] (g)
            };
        \end{feynman}
    \end{tikzpicture}
    \label{fig:Kaon2body}
    }
    \label{fig:KaonProduction}
   \captionsetup{justification=Justified,singlelinecheck=false}
    \caption{Feynman diagrams depicting the production of scalars from charged kaons. (a) Production of both HPS and flavor-specific scalars from the charged lepton leg. (b) HPS produced from the $W$ boson leg. (c) Feynman diagram for two-body decay of $K^+$ to $\pi^+$ and HPS $S$ which couples to an intermediate top quark.}
\end{figure}
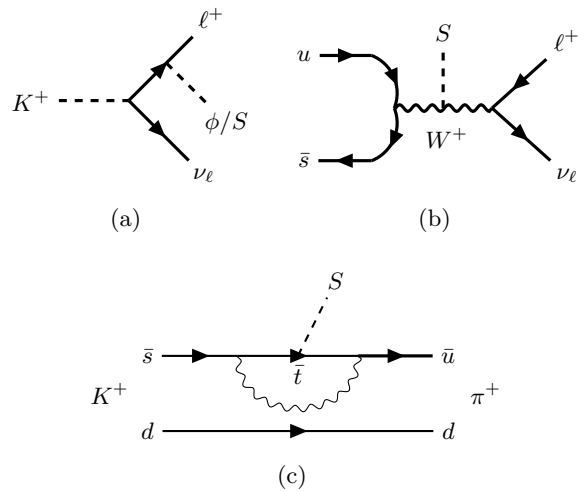
 
Charged pions and kaons dominantly decay into a charged lepton $\ell$ ($=\mu$ or $e$) and its neutrino counterpart through an off-shell intermediate $W$ boson, for example, $K^+/\pi^+ \rightarrow \mu^+ \nu_{\mu}$ and $K^+/\pi^+ \rightarrow e^+ \nu_{e}$. 
However, the above two-body decay processes are suppressed due to the required helicity of final state particles, which constrains the allowed phase space.  
This enables us to explore the production of long-lived mediators as the third decay product of charged mesons.  Unlike the corresponding two-body decay, this three-body decay would not be limited by the helicity suppression~\cite{Barger:2011mt,Laha:2013xua,rislow:3body}. 
However, the branching fraction for a choice of coupling must not exceed the upper limit on three-body decay branching fractions of charged kaons~\cite{NA62:2021bji} and charged pions~\cite{PIENU:2021clt}.
We use the three-body decay of kaons as our example in this section since the kinematically allowed phase space and mass range are larger than those of pions, although the same argument can be applied to pions.

HPS can emanate from the charged lepton leg as in Fig.~\ref{fig:leptonscalar} as well as from the $W$ boson leg~\cite{Chivukula:1988gp} as in Fig.~\ref{fig:wbosonscalar}. 
Since they couple to leptons with a strength of $m_{\ell}/v$ and $2m_{W}^2/v$ with the $W$ boson, the latter contribution dominates the relevant decay matrix element. 
The HPS can couple to charged kaons, whose strength can be calculated from chiral perturbation theory, but since this term is subdominant in the relevant decay matrix element, we do not include their contribution to the decay width. Since we take the neutrinos to be massless leptons, we omit their contribution. HPS from $K^+ \rightarrow \mu^+ \nu_{\mu} S$ are kinematically restricted to the maximum mass reach $m_K - m_{\mu} = 388$~MeV, whereas those from $K^+ \rightarrow e^+ \nu_{e} S$ can be as heavy as $m_K - m_{e}=492$~MeV.

HPS can also be produced via a kaon two-body decay, i.e., $K^+ \rightarrow \pi^+ S$ (Fig.~\ref{fig:Kaon2body}) where the scalar couples to an intermediate top quark~\cite{Bezrukov:2009yw,Gunion:1989we,BatellHPS:2019}. 
This strong coupling to the top quark makes the branching ratio of the above two-body decay process dominate over all the three-body decays, but the HPS mass is limited to $m_K - m_{\pi}=354$~MeV. 
HPS can also be produced via $B^+ \rightarrow K^+ S$, which have been searched at LHCb~\cite{LHCb:2016awg}, but the flux of $B$ mesons is not large enough at the aforementioned neutrino facilities. Hence HPS sourced from $B^+$ do not produce a sizable signal flux here.

While looking at scalars with flavor-specific couplings such as muonphilic (electrophilic) scalars, the only possible diagrams are those where scalars emerge from muon (electron) legs (Fig.~\ref{fig:leptonscalar}). Therefore, they can appear from kaons via the process $K^+ \rightarrow \mu^+ \nu_{\mu} \phi_{\mu}$ ($K^+ \rightarrow e^+ \nu_{e} \phi_e$) where the amplitude depends on Yukawa coupling squared $y_{22}^2$ ($y_{11}^2$). 

\subsection{Photons}

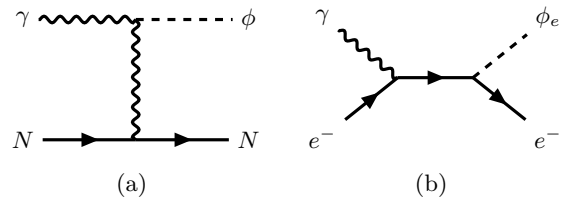
\begin{figure}[t]
    \centering
    \subfloat[]
    {
    \begin{tikzpicture}
        \begin{feynman}
            \vertex (m) at (0,0.8);
            \vertex (a) at (-1.5,0.8) {$\gamma$};
            \vertex (b) at (1.5, 0.8) {$\phi$};
            \vertex (c) at (-1.5, -0.8) {$N$};
            \vertex (d) at (1.5, -0.8) {$N$};
            \vertex (n) at (0, -0.8);
            \diagram*{
            (a) -- [boson, very thick] (m) -- [scalar, dotted, very thick] (b),
            (m) -- [boson, very thick](n),
            (c) -- [fermion, very thick](n) -- [fermion, very thick](d)
            };
        \end{feynman}
    \end{tikzpicture}
    \label{fig:scalarprim}
    }
    \vspace{0.3cm}
    \subfloat[]
    {
    \begin{tikzpicture}
        \begin{feynman}
            \vertex (m) at (-0.5,0);
            \vertex (a) at (-1.5,0.8) {$\gamma$};
            \vertex (b) at (1.5, 0.8) {$\phi_e$};
            \vertex (c) at (-1.5, -0.8) {$e^-$};
            \vertex (d) at (1.5, -0.8) {$e^-$};
            \vertex (n) at (0.5,0);
            \diagram*{
            (a) -- [boson, very thick] (m),
            (n) -- [scalar, very thick](b),
            (c) -- [fermion, very thick](m) --[fermion,very thick](n) -- [fermion, very thick](d)
            };
        \end{feynman}
    \end{tikzpicture}
    \label{fig:comptonscalar}
    }
    \captionsetup{justification=Justified, singlelinecheck=false}
    \caption{Feynman diagrams depicting the production of scalars from photons. (a) Primakoff scattering of a photon to produce HPS as well as flavor-specific scalars. (b) Compton-like scattering of a photon to produce an electrophilic scalar.}
    \label{fig:photonscalar}
\end{figure}

Scalars couple to two photons at the one-loop level. This coupling can be written as~\cite{PhysRevD.8.172, Marciano:2011gm},
\begin{equation}
    S^{\mu \nu}= g_{\phi \gamma}(p_1.p_2\eta^{\mu \nu}-p_{1}^\nu p_{2}^\mu)\,,
    \label{eq:loop}
\end{equation}

\begin{figure}[t]
    \centering
    \includegraphics[width=0.45\textwidth]{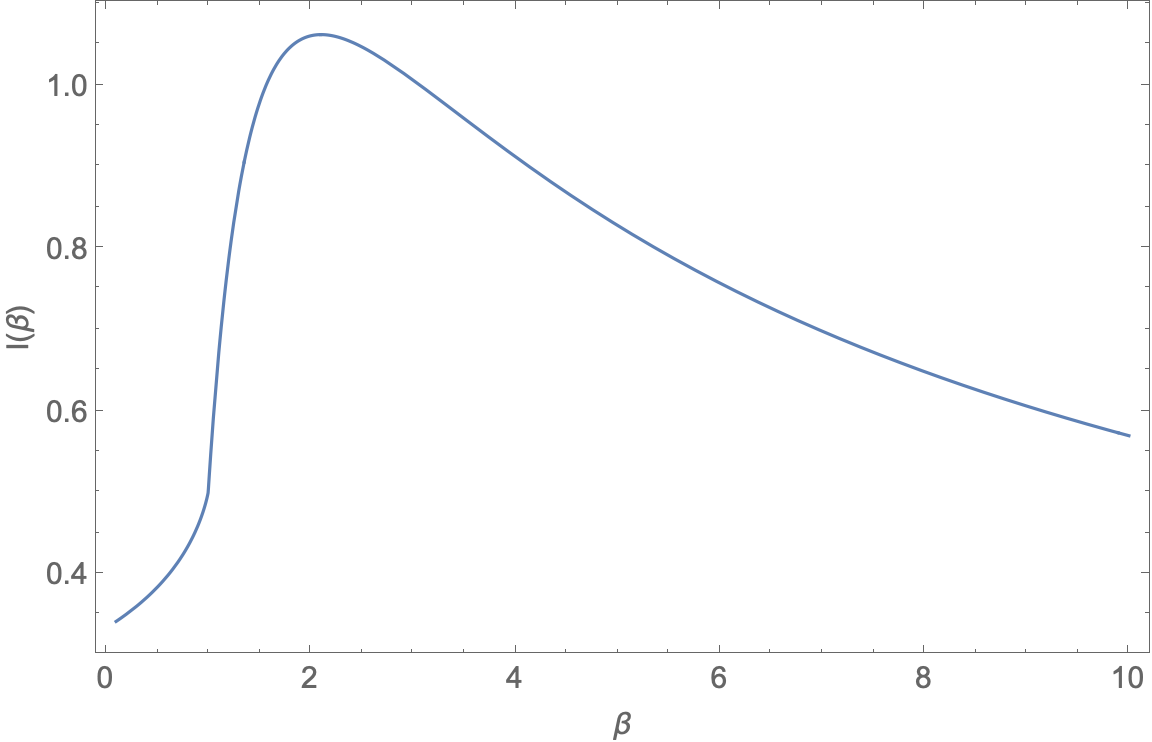}
    \caption{Variation of $I(\beta)$ with $\beta$}
    \label{fig:Ibeta}
\end{figure}

Here, the coupling strength $g_{\phi \gamma}$ is written in terms of the non-divergent one-loop factor with mass dimension $-1$\footnote{The subscript $f$ in $g_{\phi \gamma}$ is used to denote that all fermions, leptons and quarks, that couple to the scalar contribute to the loop.}:
\begin{equation}
    g_{\phi \gamma} = \frac{\alpha_{\rm em}}{\pi}\sum_{f} y_{ff} \frac{N_c Q_f^2}{m_f}I\Big(\frac{m_{\phi}^2}{4m_f^2}\Big).
    \label{eq:phigamma}
\end{equation}

where $\alpha_{\rm em} = 1/137$ is the electromagnetic fine structure constant, $p_1$ and $p_2$ denote the momenta of the two photons, and the function $I(\beta)$ carries information about the non-divergent fermion loop. It is generally expressed as
\begin{equation}
    \begin{aligned}
        I(\beta) &= \int_0^1 dx \int_0^{1-x} dy \frac{1 - 4xy}{1-4xy\beta} \\
        &= \frac{1}{2 \beta^2} [\beta + (\beta - 1) f(\beta)], 
    \end{aligned}
    \label{eq:Ifunc}
\end{equation}
where $f(\beta)$ is defined as 
\begin{equation}
    f(\beta) =
    \begin{cases}
        \arcsin(\sqrt{\beta})^2 \hspace{1cm}  & \text{for}~\beta \leq 1 \\
        -0.5 (-2 \arccosh{\sqrt{\beta}} + i\pi)^2 & \text{for}~\beta > 1 
    \end{cases}.
\end{equation}
The coupling that appears in Eq.~\eqref{eq:phigamma} is 
\begin{equation}
    \begin{aligned}
        y_{\ell\ell} = \theta m_\ell/v \hspace{0.5cm}  & \text{for}~ \text{HPS}, \\
        y_{\mu \mu} = y_{22} \hspace{0.5cm} & \text{for}~ \phi_{\mu},\\
        y_{ee} = y_{11} \hspace{0.5cm} & \text{for}~ \phi_{e}.
    \end{aligned}
    \label{eq:yll}
\end{equation}
The dot product $p_1.p_2$ equals $m_{\phi}^2/2$ if the scalar decays into two photons, and $(m_{\phi}^2 - q^2)/2$ if one of the photons is an off-shell propagator that appears in the Primakoff scattering Feynman diagram (Fig.~\ref{fig:scalarprim}). From Fig.~\ref{fig:Ibeta}, we see that the loop factor is maximized when $\beta$ is lying between 1 and 3, and it drops as $\beta \rightarrow 0, \infty$. 
For a scalar that appears in the HPS model, all massive fermions contribute to the loop, but the dominant contribution is from those fermions with masses comparable to that of the scalar, as can be seen in the argument of $I$ in Eq.~\eqref{eq:Ifunc}. However, for those that appear in the muonphilic (electrophilic) scalar model, the only contribution to the fermion loop is from muons (electrons), which are proportional to $y_{22}$ ($y_{11}$).

Through the above one-loop coupling, scalars can be produced from photons at the target via the Primakoff process, which is enhanced by a factor of $Z^2$ from the nuclear form factor as mentioned earlier. These scalars are also highly forward-directed, i.e., in the same direction as the incoming photon. Despite this enhancement, HPS produced from kaon two-body decays is more than those produced via Primakoff scattering due to the presence of $W$ boson and top quark couplings in the former scenario. Since these couplings do not exist in the case of muonphilic and electrophilic scalar models, the Primakoff production here is not as suppressed as it is in the case of HPS. 
We observe that this contribution exceeds that from kaon decays for electrophilic scalars with masses close to 1~MeV (twice the electron mass). 

Electrophilic scalars can also appear when photons interact with electrons via Compton-like scattering (Fig.~\ref{fig:comptonscalar}). Although the enhancement factor here is only proportional to $Z$, unlike the Primakoff enhancement proportional to $Z^2$, this process dominates over the scalar Primakoff process as it occurs at the tree level. For both the Primakoff and Compton-like scatterings, the minimum energy required to produce a scalar is 
\begin{equation}
    E_{\gamma} = \frac{m_{\phi}^2}{2m_T} + m_{\phi},
\end{equation}
where $m_T$ is electron mass $m_e$ for Compton-like scattering and nucleus mass $m_N$ for Primakoff scattering. 

\subsection{Electrons and positrons}

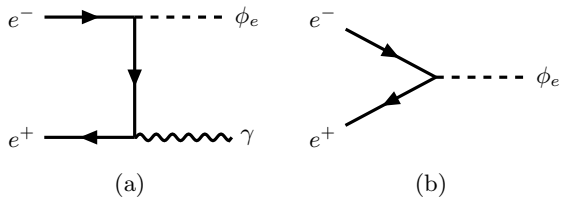
\begin{figure}[t]
    \centering
    \hspace{0.2cm}
    \subfloat[]
    {
    \begin{tikzpicture}
        \begin{feynman}
            \vertex (a) at (-1.5,0.8){$e^{-}$};
            \vertex (b) at (0,0.8);
            \vertex (e) at (0,-0.8);
            \vertex (c) at (-1.5,-0.8){$e^{+}$};
            \vertex (d) at (1.5,-0.8) {$\gamma$};
            \vertex (f) at (1.5,0.8) {$\phi_e$};
            \diagram*{
            (a) -- [fermion,very thick](b) -- [fermion,very thick](e) -- [fermion,very thick](c),
            (e) -- [photon, very thick] (d),
            (b) -- [scalar, very thick] (f)
            };
        \end{feynman}
    \end{tikzpicture}
    \label{fig:associated}
    }
    \vspace{0.5cm}
    \subfloat[]
    {
    \begin{tikzpicture}
        \begin{feynman}
            \vertex (a) at (-1.5,0.8){$e^{-}$};
            \vertex (c) at (-1.5,-0.8){$e^{+}$};
            \vertex (d) at (0,0);
            \vertex (f) at (1.5,0) {$\phi_e$};
            \diagram*{
            (a) -- [fermion,very thick](d) -- [fermion,very thick](c),
            (d) -- [scalar, very thick] (f)
            };
        \end{feynman}
    \end{tikzpicture}
    \label{fig:resonance}
    }
    \captionsetup{justification=Justified, singlelinecheck=false}
    \caption{Feynman diagrams depicting the production of scalars from electrons and positrons: (a) associated production with positrons impinging on target electrons and (b) resonance production.}
    \label{fig:enter-label}
\end{figure}

When fast-moving positrons interact with electrons at the target, they can annihilate to a photon and electrophilic scalar, $e^{+} e^{-} \rightarrow \gamma \phi_e$, as shown in Fig.~\ref{fig:associated}. However, if the energy of the incoming positron is resonated with a particular scalar mass, the electrophilic scalar can be produced directly, $e^+ e^- \rightarrow \phi_e$ (Fig.~\ref{fig:resonance}). The cross-section of this process is given by 
\begin{equation}
    \sigma_{\phi_e} = 4 \pi y_{11}^2 \frac{m_e}{m_{\phi}^2} \sqrt{m_{\phi}^2 - 4 m_e^2} \delta(E_+ + m_e - \frac{m_{\phi}^2}{2 m_e}),
\end{equation}
where $E_+$ is the energy of the incident positron.

In order to produce scalars via resonance processes, the center of mass energy of $e^+ e^-$ system must exactly match with the mass of the scalar (modulo its decay width) as suggested by the delta function. Since the center of mass energy $\sqrt{s}$ is $\sim10$~MeV for the NuMI and LBNF, where the peak energy of positrons is $\sim100$~MeV, scalars that are as heavy as 10~MeV could appear through this process. Similarly, at BNB where the peak energy of positrons is 10 MeV, a scalar of mass 3 MeV is preferred for resonant production. However, we find that this process is still subdominant at these resonant masses as compared to other processes. Therefore, we ignore the contributions from resonance while simulating scalars.

\begin{figure*}[t]
    \centering
    \begin{subfigure}[t]{.32\textwidth}
        \includegraphics[width = \textwidth]{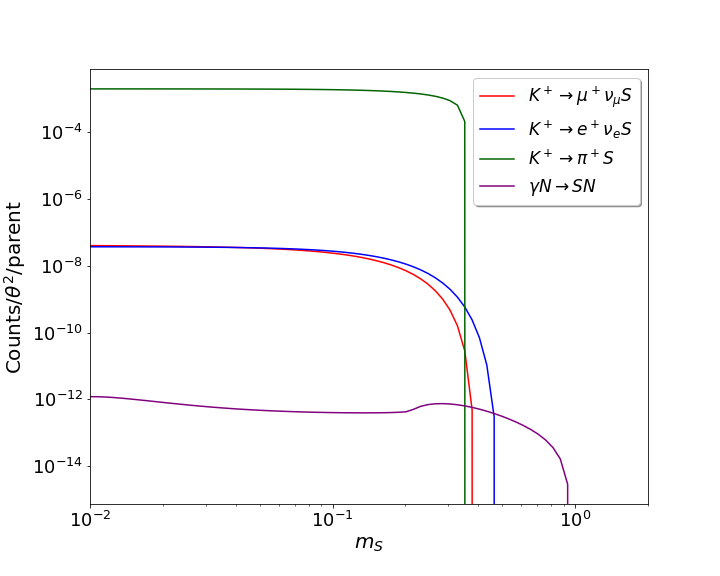}
    \end{subfigure}
    \hspace{0.1cm}
    \begin{subfigure}[t]{.32\textwidth}
        \centering
        \includegraphics[width = \textwidth]{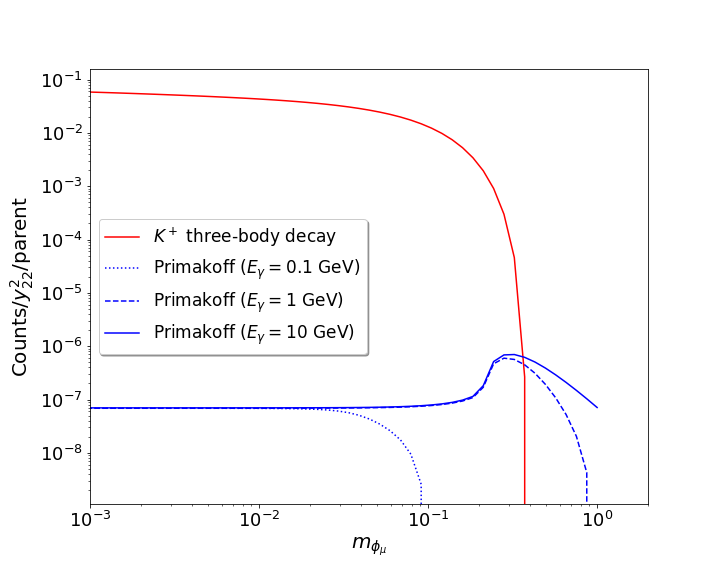}
    \end{subfigure}
    \hspace{0.1cm}
    \begin{subfigure}[t]{.32\textwidth}
        \centering
        \includegraphics[width = \textwidth]{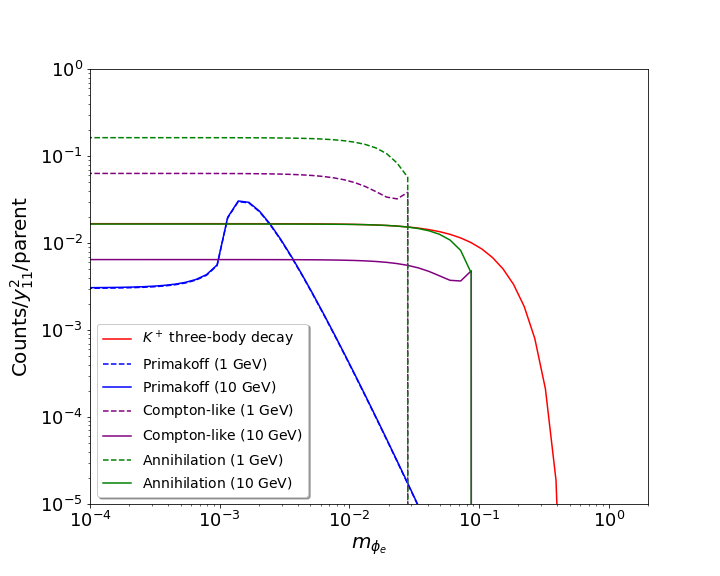}
    \end{subfigure}
    \captionsetup{justification=Justified, singlelinecheck=false}
    \caption{Number of scalars produced from each source per unit coupling. Left: Higgs Portal Scalars, Center: Muonphilic scalars, and Right: Electrophilic scalars}
    \label{fig:production}
\end{figure*}

\subsection{Simulation methods}

We use the simulated fluxes of source particles, i.e., mesons, photons, electrons, and positrons, using the \texttt{GEANT4} code package~\cite{geant4}. The flux generated is based on a 120 GeV proton beam that impinges on a 150 cm graphite target. The distribution and normalization of charged-meson fluxes as they pass by the magnet are adjusted according to magnet specifications needed for these experiments~\cite{josh,BatellHPS:2019}. We also consider the secondary production of photons and electrons. To simulate mediators, we first calculate the probability of producing them in the center-of-mass (C.O.M) frame of the production process. Using this probability distribution, we simulate mediators in the C.O.M frame and then boost them to the laboratory frame. The probability functions in the C.O.M frame are explained below:
\begin{enumerate}
    \item If a mediator is produced as a product of a two-body decay, such as $K^+ \rightarrow \pi^+ S$, the branching ratio and the energy of the mediator are fixed for a given mass of the mediator. 
    \item If produced from a three-body decay such as $K^+ \rightarrow e^+ \nu_{e} \phi_e$, we calculate the differential branching ratio as a function of energy in the rest-frame of the decaying particle. We generate random energy between the minimum and maximum energies in this frame which are weighted by the flux times the Monte Carlo volume. 
    \item If the mediator is from a 2-to-2 scattering process, for example, Compton-like scattering, we calculate $(1/{\sigma_{\rm tot}})d\sigma_{2 \to 2}/{dt}$ in the C.O.M frame of the process. In the above example, $\sigma_{\rm tot}$ is the total scattering cross section of a photon of a given energy and $\sigma_{2 \to 2}$ is the cross section of the Compton-like scattering. Also, $t$ is one of the Lorentz-invariant Mandelstam variables. In this choice of frame, the energy and momenta of all the final state particles are fixed.  
\end{enumerate}

The angular distributions of the decay processes are approximately uniform (modulo the internal propagator effect). Hence, we randomly choose the angles in the rest-frame of the decaying particle, and then boost the four-momentum along the direction of the decaying particle to the laboratory frame.
For the scattering processes, however, the angular distribution is not necessarily uniform, but a function of the Mandelstam $t$. Thus we simulate $t$'s and the appropriate Monte Carlo weights in the C.O.M frame, and then boost it back to the laboratory frame.

We record the (1) energy (2) polar and azimuthal angles, and (3) production point of the mediators in the laboratory frame for every possible mass. For each recorded event, we check whether the direction of the mediator is within the acceptance cone that the detector subtends at the point where it is produced. If so, we accept the event for the detector of interest and if not, we reject them. 

Figure~\ref{fig:production} depicts various production mechanisms that were discussed above in the context of the three models. We notice the dominance of each channel for different mass ranges for all three models. For HPS with masses less than 354~MeV, two-body decays of the kaons dominate over all the other sources by four orders of magnitude. Beyond this mass, however, the three-body decays of kaons and the Primakoff process become relevant in order. For muonphilic scalars, the three-body decays dominate for $m_{\phi} < 388$ MeV. Finally, in the case of the electrophilic scalars, the contribution from Compton-like scattering and the annihilation process are the most dominant.

While these plots help understand the probability of a scalar produced through a particular decay/scattering mechanism, we emphasize that the number of source particles and the energy spectra of both source particles and scalars are crucial when estimating the sensitivity. 
Hence, every production mechanism can be relevant and should be carefully analyzed. 
For example, electrophilic scalars from Compton-like scattering are more in number than those from $e^+ e^-$ annihilation at the aforementioned facilities since there are more photons than positrons. However, the scalars from the above two processes are softer in energy as compared to those that are produced via kaon decays. Therefore, depending on the analysis strategies (e.g., energy cut), one production channel could stand out more than the others and vice versa. 

\section{Detection of mediators}\label{Signals}

In this section, we discuss methods to detect the above-mentioned mediators at our benchmark detectors all of which adopt the liquid argon time projection chamber (LArTPC) technology. After collecting the mediators within the solid angle of the detector, they can be detected if their lifetime is long enough to survive up to the detector, without decaying into (in)visible particles before they reach the detector.  
To calculate the lifetime of the mediator for a given mass and momentum, we calculate the total decay width, which is the sum of the individual decay widths of all allowed decay channels. 

\subsection{Decay widths of mediators}

Scalars can decay into a lepton and an anti-lepton if the scalar mass is greater than twice the mass of the lepton. The decay width of a scalar with mass $m_{\phi}$ that couples to a lepton $\ell$ with strength $y_{\ell\ell}$ is
\begin{equation} \label{scalarleptondw}
    \Gamma_{\phi \ell\ell} = y_{\ell\ell}^2 \frac{m_{\phi}}{8\pi}\bigg( 1 - \frac{4m_\ell^2}{m_{\phi}^2}\bigg)^{3/2}\,,
\end{equation}
where $y_{\ell\ell}$ is given in Eq.~\eqref{eq:yll}.\footnote{Here we subscript $\ell$ to denote leptons as opposed to $f$ in Eq.~\eqref{eq:phigamma} which denotes fermions. This is because we have only leptonic final states that can appear from decays (not quarks).}
Scalars of any mass can decay into two photons with a decay width
\begin{equation}
    \Gamma_{\phi \gamma \gamma} = \frac{g_{\phi \gamma}^2m_{\phi}^3}{64\pi^3}.
\end{equation}

Here the coupling $g_{\phi \gamma}$ is given by Eq.~\eqref{eq:phigamma}. The $-1$ mass dimension in Eq.~\eqref{eq:phigamma} is manifested in the $1/v$ proportionality for HPS, $1/m_{\mu}$ for muonphilic scalars and $1/m_e$ for electrophilic scalars. Therefore, the inverse-mass scale of the photon mixing is the smallest for HPS, followed by muonphilic scalars and then electrophilic scalars. 
Additionally, HPS could also decay into two pions as well, whose decay width can be calculated from chiral perturbation theory \cite{Donoghue:1990xh}:
\begin{equation} \label{scalarleptondw}
    \Gamma_{\phi \pi \pi} = \bigg( \frac{2}{9} m_{\phi}^2 + \frac{11}{9} m_{\pi}^2\bigg)^2 \frac{3\theta^2}{32\pi v^2 m_{\phi}}\bigg( 1 - \frac{4m_{\pi}^2}{m_{\phi}^2}\bigg)^{1/2}. 
\end{equation} 
 
We will now summarize all the possible decay channels in the context of the three scalar models.
HPS decay into di-photons if their mass is less than 1 MeV.
If they are heavier than 1 MeV, the electron-positron decay channel opens up. Above 210 MeV, the muon-antimuon decay channel dominates over the electrons, and above 276 MeV, the $\pi^+ \pi^-$ decay channel adds up. If we look at the two flavor-specific scalars, muonphilic (electrophilic) scalars up to 210~MeV (1~MeV), prominently decay into two photons. However, if they are heavier than 210~MeV (1~MeV), the muon-antimuon (electron-positron) decays take over. 

\subsection{Detection channels}

For a given decay width of a mediator, the probability that it survives until it reaches the detector is 
\begin{equation}
     P_{\text{surv}} = \emph{e}^{-D/{\lambda_L}}, 
\end{equation}
where $D$ is the distance between the production point of the mediator and the front end of the detector and $\lambda_L$ is the laboratory-frame mean decay length. $\lambda_L$ can be related to the lifetime in the laboratory-frame ($t_L$) and the total decay width ($\Gamma_0$) by the following equality. 
\begin{equation}
    \begin{aligned}
        \lambda_L &= v t_L\\
        &= 0.197 \times 10^{-15}[\text{GeV}\cdot{\rm m}]\frac{p_X}{m_X}\frac{1}{\Gamma_0[\text{GeV}]},
    \end{aligned}
\end{equation}
where $m_X$ and $p_X$ are the mass and the momentum of the mediator respectively.

After reaching the detector, it can be detected if they either decay into visible particles or scatter with an argon nucleus to give rise to visible particles. 

For the decays, we require that it decays within the fiducial volume of the detector. Therefore, the probability of detecting a mediator inside a detector of length $\Delta$ is given by
\begin{equation}
    P_{\text{decay}} = \emph{e}^{-D/{\lambda_L}} ( 1 - \emph{e}^{-\Delta/{\lambda_L}}).
\end{equation}

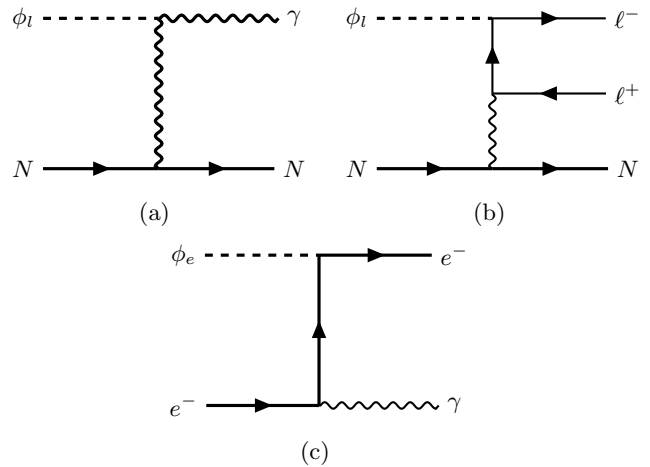
\begin{figure}[t]
    \subfloat[]
    {
    \begin{tikzpicture}
        \begin{feynman}
            \vertex (m) at (0,1);
            \vertex (a) at (-1.8,1) {$\phi_l$};
            \vertex (b) at (1.8, 1) {$\gamma$};
            \vertex (c) at (-1.8, -1) {$N$};
            \vertex (d) at (1.8, -1) {$N$};
            \vertex (n) at (0, -1);
            \diagram*{
            (a) -- [scalar, dotted, very thick] (m) -- [boson, very thick] (b),
            (m) -- [boson, very thick](n),
            (c) -- [fermion, very thick](n) -- [fermion, very thick](d)
            };
        \end{feynman}
    \end{tikzpicture}
    \label{fig:InversePrimakoff}
    }
    \subfloat[]
    {
    \begin{tikzpicture}
        \begin{feynman}
            \vertex (a) at (-1.8,1){$\phi_l$};
            \vertex (b1) at (0,1);
            \vertex (c1) at (1.8,1){$\ell^-$};
            \vertex (c2) at (1.8,0){$\ell^+$};
            \vertex (b2) at (0, 0);
            \vertex (d) at (-1.8,-1) {$N$};
            \vertex (e) at (0,-1);
            \vertex (f) at (1.8,-1) {$N$};
            \diagram*{
            (a) -- [scalar, dotted, very thick](b1),
            (c2) -- [fermion, thick] (b2) -- [fermion, thick] (b1) -- [fermion, thick](c1),
            (d) -- [fermion, very thick](e) -- [fermion, very thick](f),
            (b2) --[photon, thick](e)
            };
        \end{feynman}
    \end{tikzpicture}
    \label{fig:Splitting}
    }
    \hspace{1cm}
    \subfloat[]
    {
    \begin{tikzpicture}
        \begin{feynman}
            \vertex (a) at (-1.8,1){$\phi_e$};
            \vertex (b) at (0,1);
            \vertex (c) at (1.8,1){$e^-$};
            \vertex (d) at (-1.8,-1) {$e^-$};
            \vertex (e) at (0,-1);
            \vertex (f) at (1.8,-1) {$\gamma$};
            \diagram*{
            (a) -- [scalar, dotted, very thick](b),
            (d) -- [fermion, very thick](e) -- [fermion, very thick](b) -- [fermion, very thick](c),
            (e) --[photon, thick](f)
            };
        \end{feynman}
    \end{tikzpicture}
    \label{fig:InverseCompton}
    }
  \captionsetup{justification=Justified, singlelinecheck=off}
    \caption{Feynman diagrams illustrate the various scattering channels that mediators can undergo once they reach the detector. (a) Inverse Primakoff process where incoming scalars ($\ell = e, \mu$) scatter into a single photon. (b) Bethe-Heitler splitting process. (c) Inverse Compton-like scattering of $\phi_e$ into an electron and a photon.  }
    \label{fig:scattering}
\end{figure}

Surviving mediators can give rise to a signal by scattering off electrons, nucleons, and/or nuclei at the LArTPC detector. The various possible scattering processes of scalar mediators are: 
\begin{enumerate}
    \item Scalar inverse Primakoff: Scalars can produce a single photon signal through inverse Primakoff scattering by exchanging a photon with the nucleus. This is also a coherent process that is enhanced by $Z^2$ from the nuclear form factor (Fig.~\ref{fig:InversePrimakoff}).
    \item Bethe-Heitler process/splitting: This energy-dependent 2-to-3 scattering process gives rise to two leptons by scattering off of a nucleus. This is enhanced by the form factor and it can give rise to a lepton-antilepton signal even for mediators with masses less than twice the mass of the lepton (Fig.~\ref{fig:Splitting}). However, the mediator ($X$) requires a minimum threshold energy for this process to occur:
    \begin{equation}
        E_{X,\text{min}} = \frac{4m_\ell^2 + 4m_\ell m_N - m_X^2}{2 m_N}.
        \label{eq:EminSplitting}
    \end{equation}
    Therefore, this scattering process in energy-dependent.
    \item Inverse Compton-like scattering: This occurs when mediators scatter off an electron to produce a photon-electron signal at the detector. This channel is possible for HPS as well as electrophilic scalars (Fig.~\ref{fig:InverseCompton}).
\end{enumerate}
The Bethe-Heitler process and the inverse Compton-like scattering occur for vector mediators too. The inverse Compton-like channel can appear as a tree-level process (if the vector mediator is electrophilic) or as a one-loop process (if not electrophilic) by mixing with the SM photon. By calculating the cross-section of the above scattering processes, we can arrive at the probability of scattering $P_{\text{scat}}$ within the detector length $\Delta$. This is given by 
\begin{equation}
    P_{\text{scat}} = P_{\text{surv}} \times n_{T}\sigma_{\phi}\Delta\,,
\end{equation}
where $n_{T}$ is the number of target electrons/nucleons/ nuclei per unit volume and $\sigma_{\phi}$ is the scattering cross-section of the process of interest.
To clearly distinguish signal events from backgrounds at the detectors, various cuts are applied to the final-state particles. An example is the kinetic energy threshold for particle detection. According to MicroBooNE studies, we see that typical LArTPC detectors have kinetic energy thresholds around 20~MeV for leptons~\cite{lepcutoff} and 30~MeV for photons~\cite{gammacutoff}. After discussing our results under no background assumptions, we detail the impact of the kinetic energy threshold in the context of electrophilic scalars at DUNE ND in Sec.~\ref{sec:senscutoff}.

\section{Results}\label{SectionSensitivities} 

\begin{figure*}[t]
    \centering
    \begin{subfigure}[t]{.325\textwidth}
        \includegraphics[width = \textwidth]{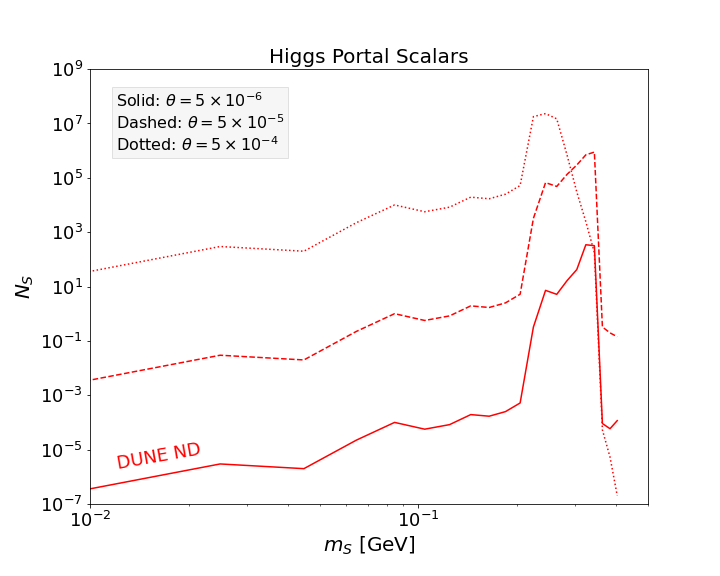}
    \end{subfigure}
    \hspace{0.01cm}
    \begin{subfigure}[t]{.325\textwidth}
        \centering
        \includegraphics[width = \textwidth]{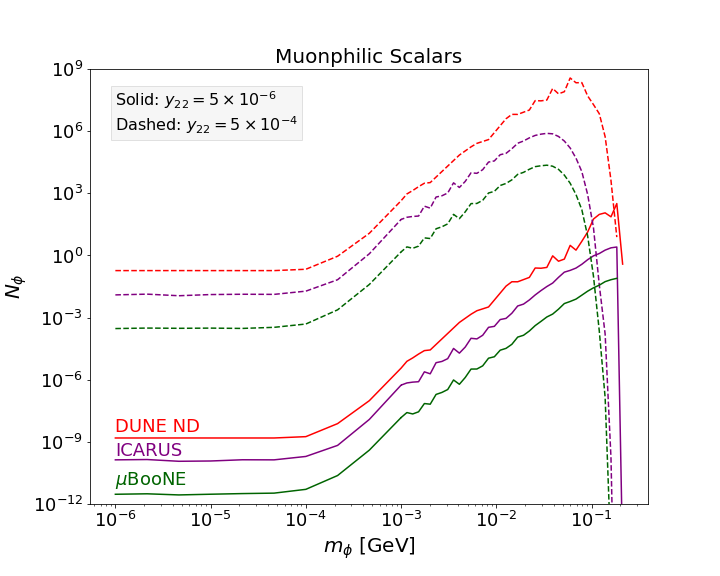}
    \end{subfigure}
    \hspace{0.01cm}
    \begin{subfigure}[t]{.325\textwidth}
        \centering
        \includegraphics[width = \textwidth]{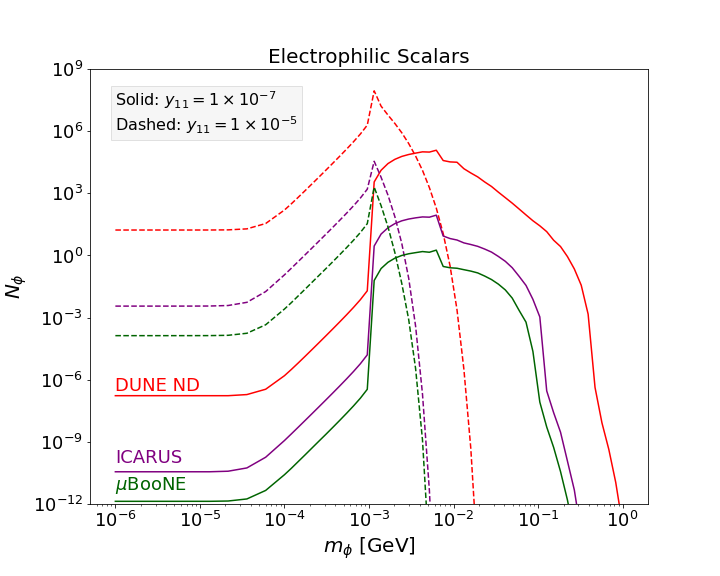}
    \end{subfigure}
    \captionsetup{justification=Justified, singlelinecheck=false}
    \caption{Number of scalars that reach the detector as a function of mass for various couplings. Left: Higgs Portal Scalars, Center: Muonphilic scalars and Right: Electrophilic scalars}
    \label{fig:events}
\end{figure*}

In this section, we report the main results of our study. Figure~\ref{fig:events} shows the number of events for different couplings as a function of mass. For the Higgs Portal Scalars, we depict the number of events at DUNE ND, whereas, for muonphilic and electrophilic scalars, we show the number of events at ICARUS and MicroBooNE as well. Since DUNE ND is placed on the beamline and is much closer to the target as compared to ICARUS and MicroBooNE to the NuMI beamline and its cumulative beam intensity is larger than that of ICARUS and MicroBooNE,
we see that, for a choice of mass and coupling, more events can be observed at DUNE ND than at the other two detectors.
ICARUS detects more events than MicroBooNE as it is not only larger but also less away from the beamline (i.e., less off beam-axis) as compared to MicroBooNE. The flat contours depict final states that appear from the scattering of the scalars in the detector whereas the linear and curved contours represent the scalar decays to photon-photon and lepton-antilepton pairs, respectively. We also notice that the number of events for larger masses [$m_{\phi} > 180$(1)~MeV for muonphilic (electrophilic) scalars] is less for larger couplings due to the shorter lifetimes. We now present the sensitivity reaches for our benchmark scalar mediator models delineated in Sec.~\ref{SectionModels}, and we also discuss the dependence of our findings on the background assumptions.

\subsection{Sensitivity estimates}

\begin{figure*}[h]
    \begin{subfigure}[t]{.48\textwidth}
        \includegraphics[width=\textwidth, clip]{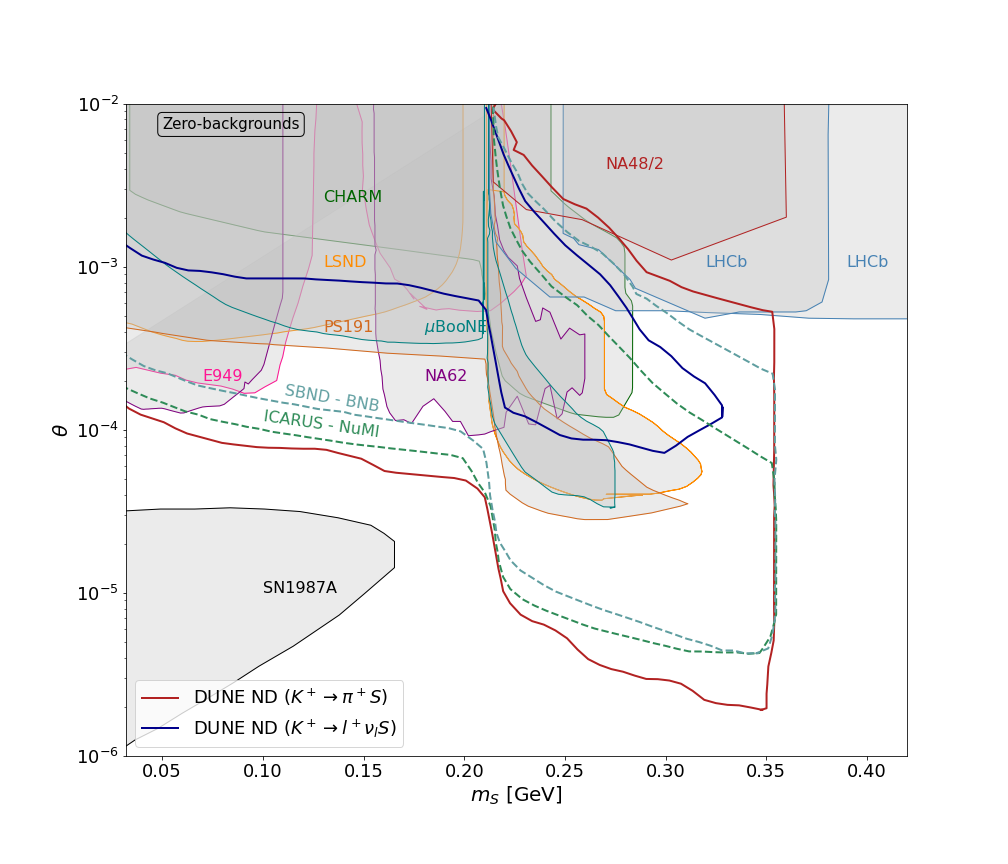}
        \captionsetup{justification=Justified, singlelinecheck=false}
        \caption{Higgs Portal Scalars. The dotted lines are from two-body decays at other experiments studied in~\cite{BatellHPS:2019}. The $95\%$ CL sensitivity lines for $m_S < 0.21~\text{GeV}$ correspond to the decay of scalars to $e^+, e^-$ only. Beyond $0.21~\text{GeV}$, the hanging lobe includes $\mu^+, \mu^-$ signals and $\pi^+, \pi^-$ as well for $m_S > 0.278~\text{GeV}$. In this plot, we include $90\%$ CL bounds imposed by CHARM~\cite{Winkler:2018qyg}, LSND~\cite{Foroughi-Abari:2020gju}, PS191~\cite{Gorbunov:2021ccu}, NA62~\cite{NA62:2021zjw}, E949~\cite{BNL-E949:2009dza}, and $95\%$ CL bounds from $\mu$BooNE~\cite{MicroBooNE:2021usw}, and LHCb~\cite{LHCb:2015nkv}.}
        \label{fig:HPS}
    \end{subfigure}
    \hspace{0.2cm}
    \begin{subfigure}[t]{.48\textwidth}
        \centering
        \includegraphics[width=\textwidth,clip]{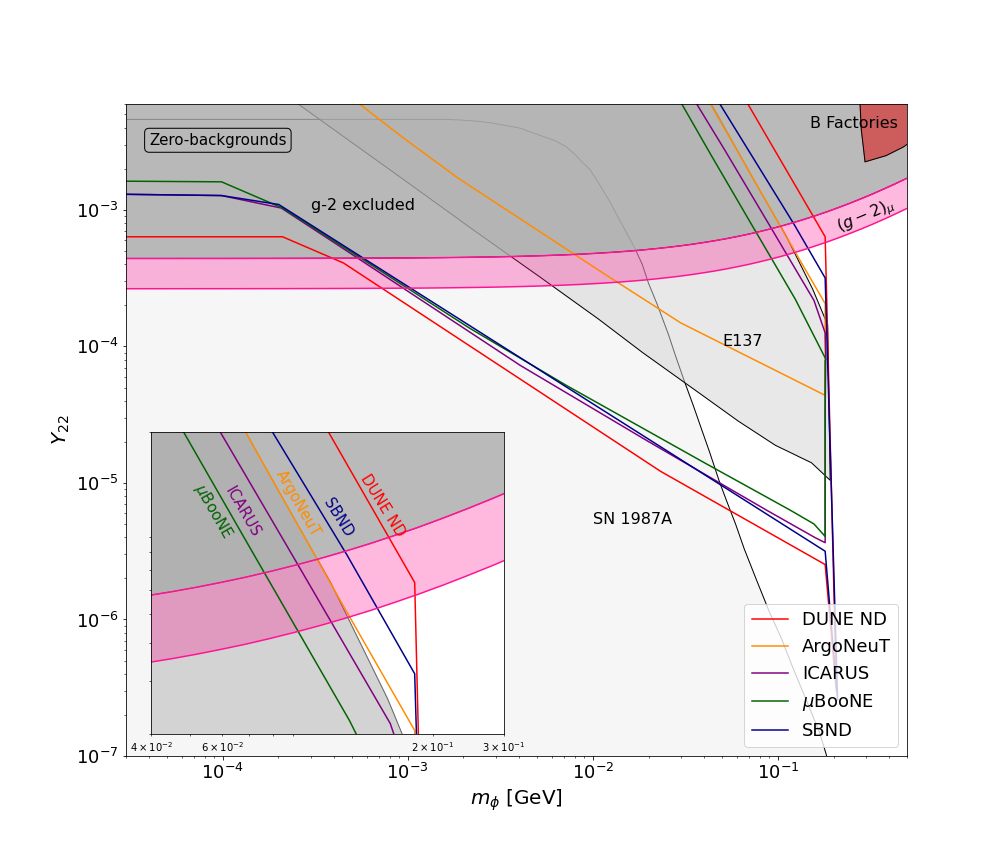}
        \captionsetup{justification=Justified, singlelinecheck=false}
        \caption{Muonphilic scalar model. The pink band shows the preferred regions of the current muon $g-2\mu$ anomaly. Scalars lighter than $0.21~\text{GeV}$ dominantly scatter via spitting into $\mu^+ \mu^-$, and decay into two photons, the latter dominating for $m_{\phi} > 0.0001~\text{GeV}$. For scalars greater than $0.21~\text{GeV}$ they decay to $\mu^+ \mu^-$, which is manifested by the sharp drop in sensitivity. The $90\%$ CL bounds from E137~\cite{Dobrich:2015jyk}, $95\%$ CL bounds from B factories, $g-2$ excluded regions, and constraints from SN cooling are depicted in the shaded regions.}
        \label{fig:muon-philic}
    \end{subfigure}\\
    \begin{subfigure}[t]{.48\textwidth}
        \centering
        \includegraphics[width=\textwidth,clip]{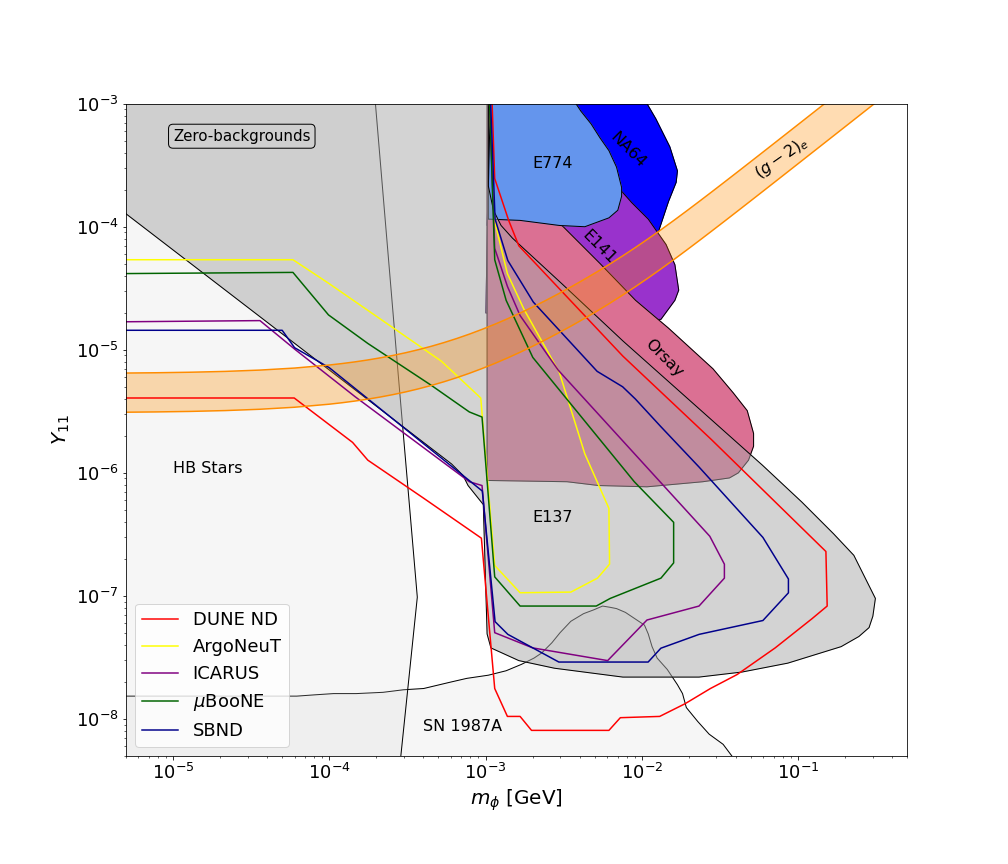}
        \captionsetup{justification=Justified, singlelinecheck=false}
        \caption{Electrophilic model. The parameters in the orange band of this model explain current electron $g-2$ uncertainty. These scalars give rise to two photons via decay and $e^+, e^-$ via splitting for $m_{\phi} < 0.001 ~\text{GeV}$, and  $e^+, e^-$ only via decay for $m_{\phi} > 0.001~\text{GeV}$. The shaded regions represent $90\%$ CL bounds from E137~\cite{Liu:2017htz}, E141~\cite{Riordan:1987aw}, E774~\cite{Bross:1989mp}, Orsay~\cite{Bechis:1979kp}, NA64~\cite{NA64:2021aiq}, and HB stars~\cite{Hardy:2016kme} and SN cooling~\cite{Lucente:2021hbp}. }
        \label{fig:electron-philic}
    \end{subfigure}   
    \caption{ $95 \%$ CL lines for our three benchmark scalar mediator models under the assumption of zero backgrounds. } 
    \label{fig:sensitivities}
\end{figure*}

The sensitivities discussed in this section are obtained under zero background assumptions. Therefore, the contour for each experiment has been plotted for 3 events, which is the upper bound of the 95\% confidence level (CL) interval for zero backgrounds. Any parameter that is enclosed inside the contours for each experiment yields more than 3 events. This is an all-inclusive sensitivity plot where the sensitivities include all possible events from decays and scatters of the mediators. Fig.~\ref{fig:sensitivities} has the three sensitivity plots for the three models. The nature of the signals induced is explained in the caption below.
\begin{enumerate}
    \item \textbf{Higgs Portal Scalars.} Figure~\ref{fig:HPS} shows the sensitivity plot of HPS. The constraints are from recent studies such as Refs.~\cite{Bhanderi:2023lwb, Batell:2023mdn, Antel:2023hkf}. We included the limits for DUNE ND while the lines for other experiments have been taken from Ref.~\cite{BatellHPS:2019}. To verify our results, we reproduced the ICARUS sensitivity given in Ref.~\cite{BatellHPS:2019}. We first note that our predictions for DUNE ND differ from those in~\cite{Antel:2023hkf, Berryman:2019dme}. These are rooted in differences in the number of events that define the sensitivity, and also energy/angular separation thresholds. The contribution to the scalars from two-body decays has been considered in Ref.~\cite{Berryman:2019dme}. Here, however, we plot the contributions of scalars produced via two-body decays (red line) and three-body decays (blue line) separately to compare. We observe that the majority of scalars are produced from the two-body decay process, $K^+ \rightarrow \pi^+ S$. This is due to the top quark coupling. However, only those scalars lighter than 354~MeV can be produced via this process. Although HPS with masses greater than 354 MeV can be produced via three-body decays,  $K^+ \rightarrow e^+ \nu_e S$ and $K^+ \rightarrow \mu^+ \nu_{\mu} S$, the flux of these scalars is suppressed. This is not only because it is a three-body decay, but also because the couplings are weaker than the previously mentioned top quark coupling.
    
    \hspace{0.2cm} We find that the sensitivity at DUNE ND is more enhanced in comparison to ICARUS, especially for larger couplings. For the masses and couplings of our interest, we see the number of scalars produced from Primakoff processes is subdominant. They are prominent only for $\theta$ values greater than $10^{-3}$, which is constrained by LHCb. For the masses and couplings of our interest, all signals produced by these scalars are from decay processes. We do not see signals from scattering processes because they are subdominant for weak couplings ($\theta < 10^{-2}$).

    \item \textbf{Muonphilic scalars.} Figure~\ref{fig:muon-philic} depicts the sensitivity plot of the muonphilic scalar model. Since these scalars do not couple to quarks, $W$ gauge bosons, or first- and third-generation leptons, their production modes are limited to kaon three-body decays $K^+ \rightarrow \mu^+ \nu_{\mu} \phi_\mu$ by coupling to the muon, and Primakoff process $\gamma N \rightarrow \phi_\mu N$. In the region represented by the current $g-2$ discrepancy (represented by the pink band), the detected signals in this region are mostly di-photons. We also see a dip in the sensitivity plot at 210~MeV where muon-antimuon decays start to appear, thus reducing the scalar lifetimes. Stringent constraints appear from the 20~GeV electron beam experiment E137~\cite{Dobrich:2015jyk}. We converted the limits on the scalar-photon coupling in ~\cite{Dobrich:2015jyk} into limits in the $y_{22}$ space and found that our results match with those in~\cite{Harris:2022vnx}. Since forward detectors such as DUNE ND, SBND, and ArgoNeuT (existing data) are more sensitive to high-energy mediators (which have longer lifetimes), the ceiling of the sensitivity curve for the above forward detectors is higher than off-axis detectors. Thus, sensitivities of forward detectors extend beyond E137 bounds, and into the $g-2$ band. We also notice that big detectors are exposed to a great intensity of muonphilic scalars. Hence, they are sensitive to a large region of parameter space which is allowed by the E137. Since ArgoNeuT is smaller in size, it is challenging for them to probe couplings smaller than those excluded by E137.
    
    \hspace{0.2cm} Some regions of this allowed parameter space that are explored by the neutrino experiments are ruled out by SN1987a data. However, the astrophysical bounds can be avoided in light mediators models by the chameleon effect~\cite{Khoury:2003aq,Brax:2007ak}, where the masses of the mediators can depend on the background matter density. It must be noted that the environment in the core of a supernova (or a star) is extremely dense. In such a highly dense environment, the effective mass of the particle can be larger than it is on Earth causing an expansion of the allowed parameter space. In any case, it is important to probe the parameter space using laboratory experiments.
    
    \hspace{0.2cm} Unlike HPS, scattering channels are relevant to this model. Amongst the two scattering channels in this model, inverse Primakoff and Bethe-Heitler splitting, the latter dominates over the former despite the phase space suppression. This is because the Bethe-Heitler process occurs at the tree level. Additionally, two Feynman diagrams contribute to the matrix element of the Bethe-Heitler process. 
    
    \item \textbf{Electrophilic scalars.} The sensitivity of this model is depicted in Fig.~\ref{fig:electron-philic}. The coupling of these scalars to electrons opens up many other production channels such as Compton-like scattering and associated production. While the most energetic scalars appear from $K^+$ decays, Compton-like scattering of photons with target electrons contributed to the most number of scalars. Since the sensitivity is majorly driven by Compton-like scattering, -- in which the cross-section depends on the energy flux of the photons, ultimately resulting in $\sigma_{\text{Compton}} \propto y_{22}^2/m_{\phi}^2$ -- we see that the floor of the sensitivity plots derived from our studies varies as compared to that in the E137 bounds where the scalars are sourced by the decay of charged mesons (i.e., their branching ratio to a scalar is proportional to $y_{22}^2$ for masses lighter than the decaying meson). This results in different shapes of the floor which depicts the difference in the form of the branching ratios. Primakoff scattering majorly contributes to scalars with mass 1~MeV. The availability of multiple sources of electrophilic scalars comes at the cost of many constraints, with E137~\cite{Liu:2017htz} being the most stringent one. However, we find that the DUNE ND is sensitive to parameters that are still unconstrained by HB Stars~\cite{Hardy:2016kme} and E137. 
    
    \hspace{0.2cm}These scalars are detected via electron-positron decay pairs for masses greater than 1~MeV, di-photons decays for masses between (0.01-1)~MeV, and through scattering channels for masses below 0.01~MeV. There are three possible scattering channels for electrophilic scalars, inverse Compton-like, splitting contribution, and inverse Primakoff. Out of these three, the splitting contribution is the most dominant scattering channel for all detectors because the minimum threshold required for electron-positron splitting is very low, much lower than the scale of NuMI and BNB.

\end{enumerate}

The splitting process, as seen in Fig.~\ref{fig:Splitting}, is unique because it results in lepton-antilepton final states even if the mediator mass is less than twice the lepton mass, which would not be kinematically allowed to decay. They become subdominant for higher mediator masses (masses close to twice the lepton mass) like all other scattering channels. However, they give us a unique way of identifying photon-less, purely leptonic final states. 

\subsection{Sensitivities with detection constraints} \label{sec:senscutoff}

In our analysis and resulting sensitivities reported in the previous section, we envisioned the situation where associated backgrounds are sufficiently suppressed. 
While careful background estimates would lead to more precise sensitivity estimates, they certainly depend on signal channels and detector capabilities such as energy threshold, energy/angular resolutions, and particle identification.
In particular, since the LArTPC technology is being developed and matured, higher-capability detectors would allow for rejecting more backgrounds while retaining signals.
Nevertheless, in this section, we investigate how our sensitivity results are affected by the background assumption, especially in the context of DUNE ND. 
\begin{figure*}[h]
    \begin{subfigure}[t]{.48\textwidth}
        \includegraphics[width = \textwidth]{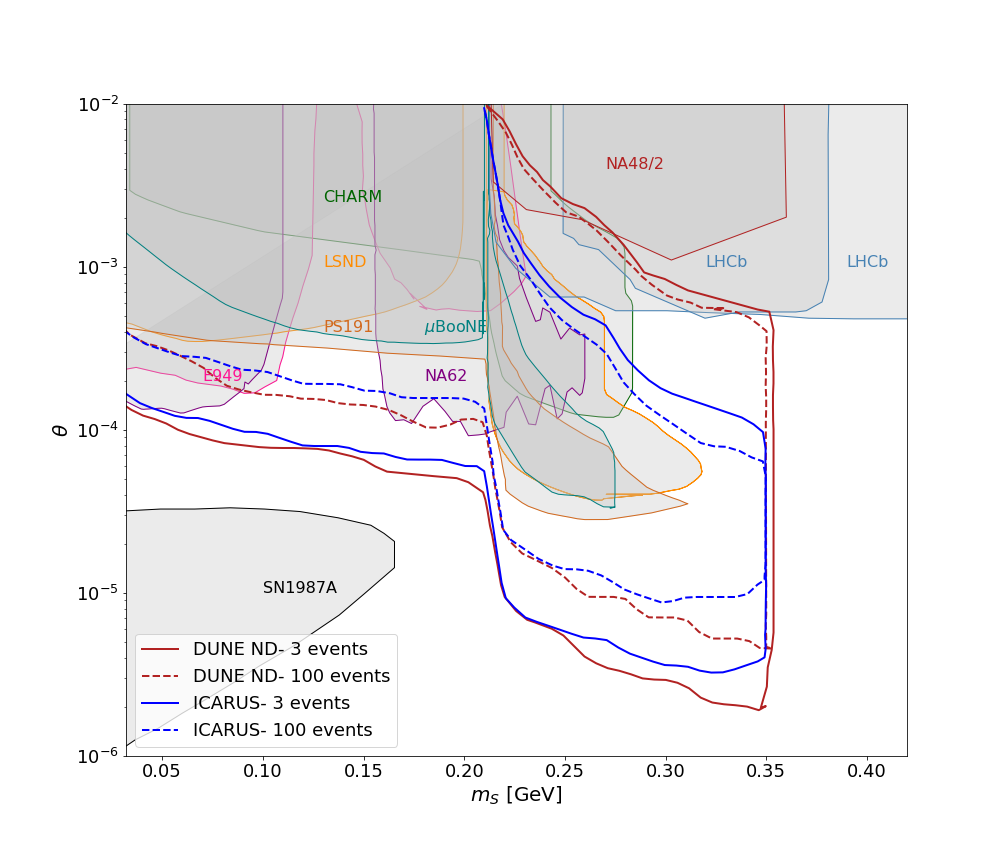}
        \captionsetup{justification=Justified, singlelinecheck=false}
        \caption{Sensitivity plot for HPS for 3 events as well as 100 events for DUNE ND and ICARUS. In this plot, we include $90\%$ CL bounds imposed by CHARM~\cite{Winkler:2018qyg}, LSND~\cite{Foroughi-Abari:2020gju}, PS191~\cite{Gorbunov:2021ccu}, NA62~\cite{NA62:2021zjw}, E949~\cite{BNL-E949:2009dza}, and $95\%$ CL bounds from $\mu$BooNE~\cite{MicroBooNE:2021usw}, and LHCb~\cite{LHCb:2015nkv}.}
        \label{fig:hpsbkg}
    \end{subfigure}
    \hspace{0.2cm}
    \begin{subfigure}[t]{.48\textwidth}
        \centering
        \includegraphics[width = \textwidth]{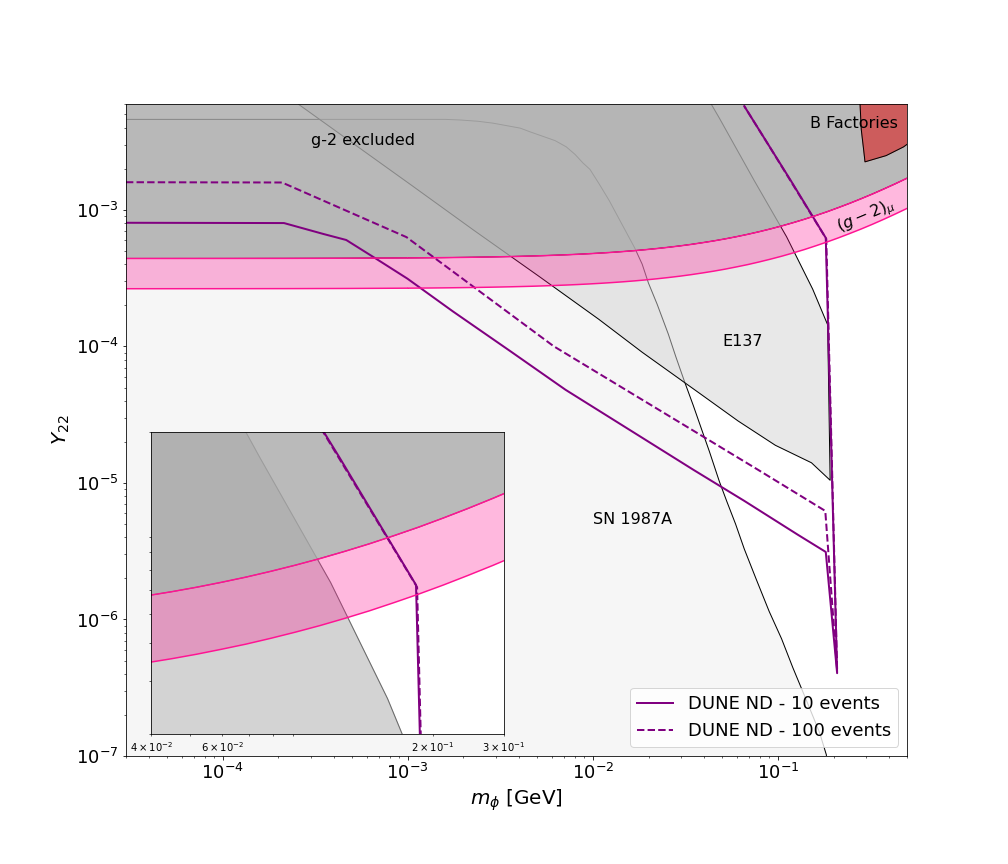}
        \captionsetup{justification=Justified, singlelinecheck=false}
        \caption{Parameters for 10, 100 muonphilic scalar decays at DUNE ND. The $90\%$ CL bounds from E137~\cite{Dobrich:2015jyk}, $95\%$ CL bounds from B factories, $g-2$ excluded regions, and constraints from SN cooling are depicted in the shaded regions.}
        \label{fig:muonscalarDUNE}
    \end{subfigure}\\
    \begin{subfigure}[t]{.48\textwidth}
        \centering
        \includegraphics[width = \textwidth]{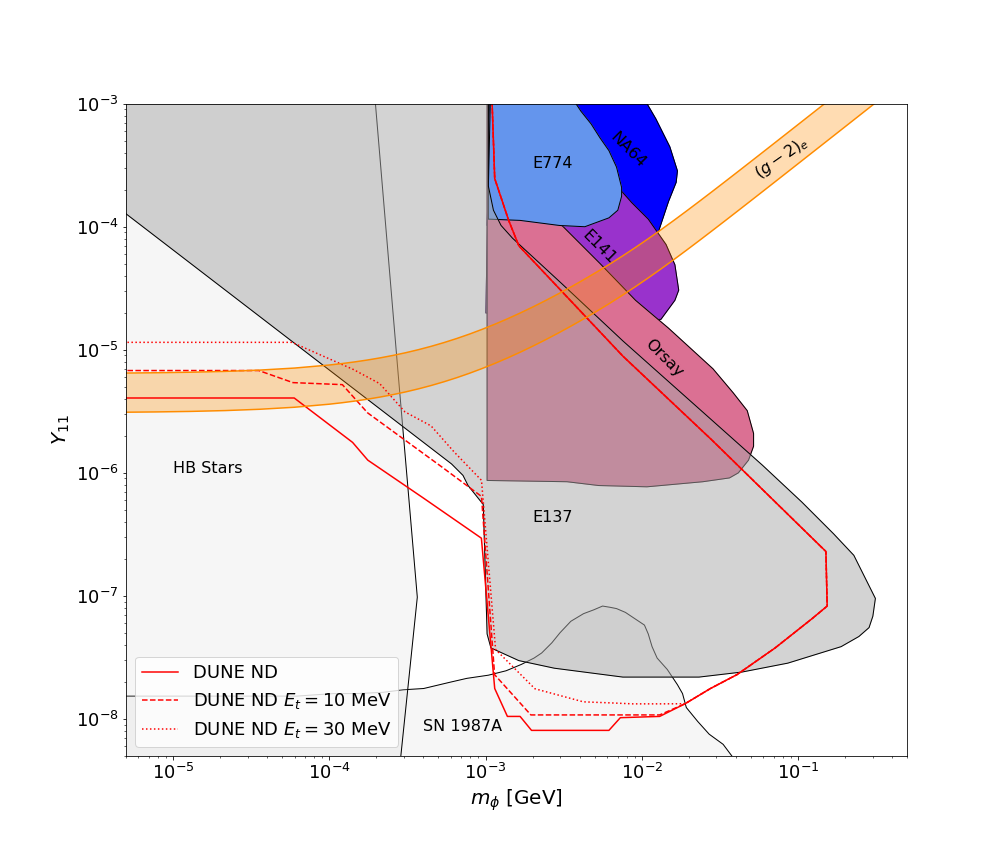}
        \captionsetup{justification=Justified, singlelinecheck=false}
        \caption{Sensitivity plots for electrophilic scalars at DUNE ND where the minimum threshold kinetic energy ($E_t$) required to identify the decay products is 0~MeV, 10~MeV, and 30~MeV. The shaded regions represent $90\%$ CL bounds from E137~\cite{Liu:2017htz}, E141~\cite{Riordan:1987aw}, E774~\cite{Bross:1989mp}, Orsay~\cite{Bechis:1979kp}, NA64~\cite{NA64:2021aiq}, and HB stars~\cite{Hardy:2016kme} and SN cooling~\cite{Lucente:2021hbp}.}
        \label{fig:y11bkg}
    \end{subfigure}   
    \caption{ Sensitivities where experimental and detection constraints are considered. } 
    \label{fig:sensitivitiesbkg}
\end{figure*}

The backgrounds for the HPS model at ICARUS and SBND have been investigated in Ref.~\cite{BatellHPS:2019}. Figure~\ref{fig:hpsbkg} roughly demonstrates the effect of backgrounds at DUNE ND by plotting the sensitivity contours for 100 events along with 3 events. We plot the 100 events line as it roughly corresponds to the worst-case reduction in sensitivity seen in Fig.~15 of Ref.~\cite{BatellHPS:2019}. Though the parameter space coverage does not reduce drastically with increasing the number of signal events, improved background analysis is expected to minimize that reduction in the future. We see that the forward DUNE ND can continue to probe those parameters with short lifetimes even if a larger number of events are required to determine the sensitivity, i.e., the expected sensitivity reaches are not very sensitive to the underlying background assumption. 

Since mediators with strong couplings tend to decay rapidly, the range of couplings in the ceiling of a sensitivity lays an upper bound to the coupling strength such that 3 of them make it to the detector. This limit depends on the mass and momentum of the mediator, and the distance between the source and detector~\cite{Dutta:2023abe}. The lower edge, on the other hand, depends on statistics of mediators produced at the target which approximately scales with the square of the coupling~\cite{Dutta:2023abe}. This depends on scalings, such as the number of POT, the branching ratio of the production process, etc. If we increase the required statistics by looking for the number of events greater than 3, the minimum coupling required would increase as compared to the case of 3 events, therefore pushing up the lower edge of the sensitivity. Figure~\ref{fig:muonscalarDUNE} is an example that shows the sensitivity contours of the muonphilic scalar model at DUNE ND for 10 events and 100 events. It clearly shows that the contours do not shrink much as we increase the number of considered events. The change in number affects the lower edge of the sensitivity plot only. Since we would look for more statistics in the presence of backgrounds, they would affect the lower limits rather than the ceiling. This also implies that the presence of backgrounds would not compromise the ability of these experiments to probe parameters in the $g-2$ band, as seen in the example plot.

The minimum threshold kinetic energy ($E_t$) required to identify signals at the detector plays a vital role in arriving at sensitivity plots of certain models at certain detectors. We notice that this constraint does not reduce the extent of the sensitivity curves of the muonphilic scalar model as the majority of the mediators are produced from processes that favor high-energy mediators. However, we see that this plays a role in the sensitivity of the electrophilic scalar model, especially for masses in the keV range as scalars produced from Compton-like scattering, being the dominant one, are lower in energy, and so are the final states. Since the sensitivity estimates for DUNE ND give us the most insight into unexplored parameter spaces, we investigate the effect of the energy thresholds at the DUNE ND in this study. Figure~\ref{fig:y11bkg} demonstrates the effects of this by showing the contours without cuts versus those with 10~MeV and 30~MeV cuts. As expected, we see the reduction in sensitivity space coverage by a few factors toward the lower-mass regime. Similar to kinetic energy thresholds, the imposition of a minimum angular separation between the decay products can play a vital role in the sensitivity plots as they affect the ceiling of sensitivity reaches. We notice that the reduction factor of coupling values in the ceiling can vary from $O(10)$ to $O(1)$ based on the threshold value and the mass of the mediator. 

In summary, Fig.~\ref{fig:sensitivitiesbkg} demonstrates multiple ways in which sensitivities are altered with experimental constraints. A more detailed analysis of detector responses and backgrounds for LArTPC-type detectors will certainly allow for more precise sensitivity estimates.

\section{Vector Mediators \label{sec:vectors}}
As discussed earlier, the above analysis is not limited only to scalar mediators. They can be extended to spin-1 mediator models (gauge bosons) as well. They can be greatly produced from charged mesons present at the target. As an example, let us consider the $U(1)_{T_{3R}}$ model that appears in the context of left-right symmetric models~\cite{PhysRevD.11.703.2, PhysRevD.11.2558}. This model contains a gauge boson, $Z'$, that couples to the right-handed fermions of one particular generation. Although there exist new fields in this anomaly-free model which is a spontaneously broken model at a low energy scale, e.g., low mass scalars and dark-matter candidates, we consider the effects of the gauge boson with visible decay modes only where the gauge boson couples to the right-handed muon, charm, and strange quarks. 

These gauge bosons can be heavily sourced from three-body decays of charged mesons $K^+ \rightarrow \mu^+ \nu_{\mu} Z'$, emanating from the muon leg $\mu^+$. Since these gauge bosons couple only to the right-handed component of the muon, the muon's helicity must be flipped, thereby suppressing this three-body decay by the muon mass. Despite this condition, we observe an enhancement effect, similar to the scalar three-body decay case. This is due to the existence of the longitudinal polarization mode of vector gauge bosons that allows for an enhancement proportional to $1/m_{Z'}^2$~\cite{rislow:3body}. 
Since these gauge bosons can mix with the SM photon kinetically, they can couple to electrons with strength $\epsilon e$ where 
\begin{equation}
    \epsilon = g_{T3R}\sqrt{\frac{\alpha_{\rm em}}{{4\pi^3}}}.
\end{equation}
\hspace{0.2cm} This expands the horizon of gauge boson production including neutral meson decays, $\pi^0/\eta \rightarrow \gamma Z'$. Similarly, gauge bosons can also be produced via Compton-like scattering, $\gamma e^- \rightarrow Z' e^-$~\cite{Chakraborti:2021hfm} when photons hit the target electrons. The production mechanisms of electrophilic scalars can be applied to $Z'$ gauge bosons as well, i.e., pair annihilation $e^{+} e^{-} \rightarrow \gamma Z'$~\cite{Nardi:2018cxi} and resonance production $e^+ e^- \rightarrow Z'$~\cite{Nardi:2018cxi}. Additionally, $Z'$s can appear from electron/positron bremsstrahlung $e^{\pm} N \rightarrow e^{\pm} N Z'$~\cite{Nardi:2018cxi,Dutta:2020vop}. However, it is important to note that these processes occur through $\gamma-Z'$ mixing, which is of the order $O(10^{-2})$. Therefore, the flux of gauge bosons from three-body decays of charged mesons is much larger as they are produced via the direct coupling to the lepton. It is important to note that the three-body decay must satisfy the upper limit of the charged kaon/pion branching ratio.

We probe the sensitivity of the $U(1)_{T_{3R}}$ gauge boson through visible decays into electrons and positrons via the kinetic mixing loop. This decay width is given by 
\begin{equation}
    \Gamma_{Z',e^+e^-} = \frac{\epsilon^2 \alpha_{\rm em} m_{Z'}}{3}\bigg( 1 + \frac{2m_e^2}{m_{Z'}^2}\bigg) \sqrt{1 - \frac{4m_e^2}{m_{Z'}^2}}.
\end{equation}

We see that the production of these gauge bosons from charged mesons dominates over neutral mesons despite constraints on the three-body branching ratio. The magnetic focusing horn system facilitates this production mechanism and hence, we obtain a great amount of sensitivity for weak couplings. Due to constraints on the upper limit of the three-body branching fraction, we are unable to excavate higher couplings through this production mechanism, which are mostly constrained by experiments such as NA64, E774, etc. Thus, in this $U(1)_{T3R}$ model, we see that the production of these gauge bosons from charged mesons can give us insights into the lower coupling range that are unexplored parameters, as supported by our sensitivity study under the assumption of negligible backgrounds (see Fig.~\ref{fig:enter-label}). Since the branching ratio of a charged meson into a gauge boson is proportional to $g_{T3R}^2/m_{Z'}^2$, the floor of the sensitivity has a linear nature, with a positive slope. This leads to a unique shape of the sensitivity as compared to those in the excluded regions in which the gauge bosons are produced through two body decays of neutral mesons (where the branching ratio is roughly proportional to $g_{T3R}^2$).

\begin{figure}[t]
    \centering
    \includegraphics[width = 0.5\textwidth]{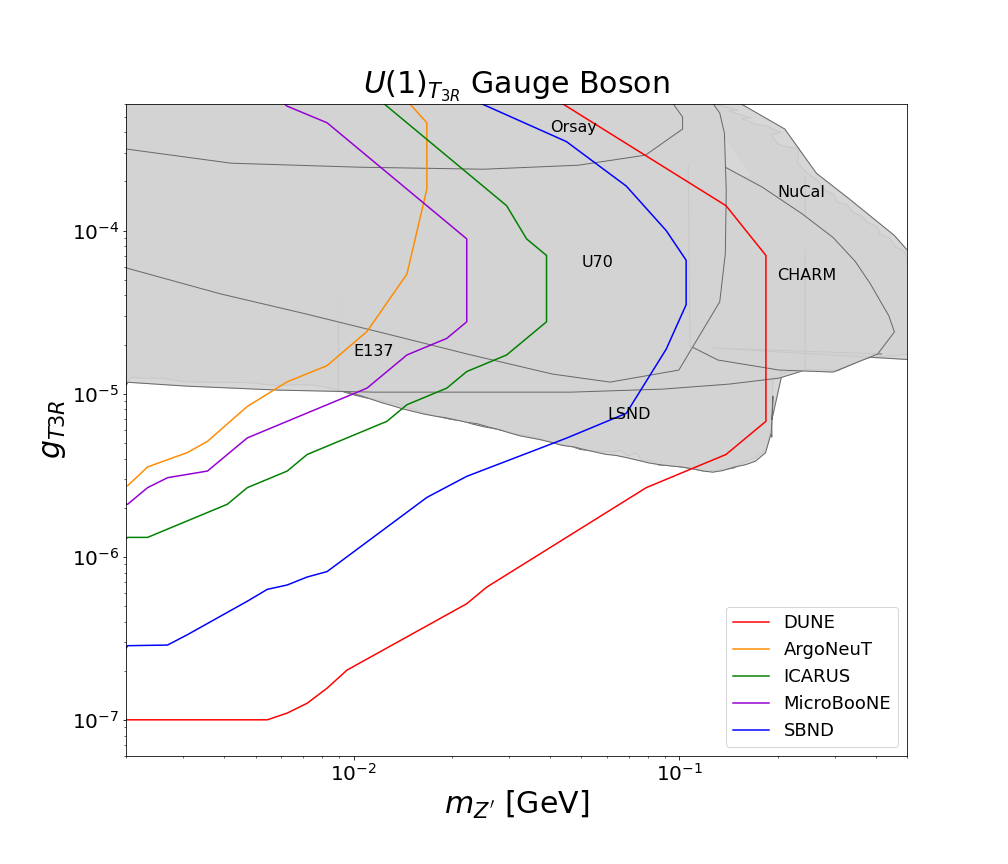}
    \captionsetup{justification=Justified, singlelinecheck=false}
    \caption{ 95\% CL sensitivity plot (with no backgrounds) of the $U(1)_{T_{3R}}$ gauge boson with only visible decays.}
    \label{fig:enter-label}
\end{figure}

Finally, we attempt to apply these production and detection mechanisms to $U(1)_{L_i-L_{j}}$ gauge bosons. Since these gauge bosons couple to neutrinos as well, their lifetimes at MeV-to-sub-GeV facilities are not long enough to suggest considerable value-added insights into unconstrained parameters.

\section{Conclusions \label{sec:conclusions}}

In this study, we explore three dark-sector models, Higgs Portal Scalars, muonphilic scalar models, and electrophilic scalar models, at neutrino experiments by utilizing the vast flux of mesons, photons, and electrons that are produced at the neutrino target. The neutrino experiments considered in this study are the finished ArgoNeuT, ongoing MicroBooNE, SBND, ICARUS, and the upcoming DUNE experiments. We have also demonstrated an example of a vector muonphilic mediator, which is the gauge boson in the $U(1)_{T3R}$ model.
The magnetic horns at the targets of these experiments have been utilized to maximize the production of the mediators from charged mesons along the beam direction. The choice of models and mediators enables us to understand the domination of one production mode over the other. Although three-body decays of mesons dominate for models with muonic couplings, and two-body meson decays dominate the HPS model, Compton-like scattering and Primakoff production from photons are most dominant for models with electron couplings. Through these production mechanisms, the potential for high-energy mediators to reach the detectors increases, especially at forward detectors. This allows us to explore the $g-2$ regions along with large regions of unexplored parameter space in the laboratory experiments at the ongoing/upcoming facilities, especially for the muon. 

Since these mediators produce visible signals in the detector through multiple scattering and decay mechanisms, we were able to get an all-inclusive sensitivity plot demonstrating the potential of probing regions of parameter space that are still unexplored by experiments. The inclusion of the energy-dependent Bethe-Heitler scattering process and the decay process allowed us to access unique lepton-antilepton signatures for flavor-specific mediators. 

A rigorous background analysis would allow us to estimate more realistic exclusion limits. A better understanding of the background would include angular separation capabilities along with exact energy thresholds. For example, Ref.~\cite{MicroBooNE:2021usw} has done a study for electron-positron events at MicroBooNE from which we find an estimate of around 30 background events. However, it is important to note that the ceiling of sensitivities does not change appreciably, as demonstrated in Fig.~\ref{fig:sensitivitiesbkg}, since they are obtained from extremely high-energy mediators. The number of events calculated in this study may have uncertainties that vary between $0.3-3$, due to the charged meson fluxes. This can vary the new physics couplings in the sensitivity estimates (Figs.~\ref{fig:sensitivities} and~\ref{fig:sensitivitiesbkg}) by factors ranging from 0.7 to 2. However, these effects are visible only on the floor of the sensitivity plots.

Our studies based on the example models can straightforwardly be extended to many more scenarios with bosonic mediators. 

\section*{Acknowledgements}
We thank Wooyoung Jang for his work on the \texttt{GEANT4} simulations. 
We would also like to thank Joshua Berger, Nityasa Mishra, Ornella Palamara, and Adrian Thompson for their useful discussions.
This work of BD, AK, and DK is supported by the U.S. Department of Energy Grant DE-SC0010813.

\appendix

\section{Parent particle fluxes}
\begin{figure}[h]
    \centering
    \includegraphics[width = 0.49\textwidth]{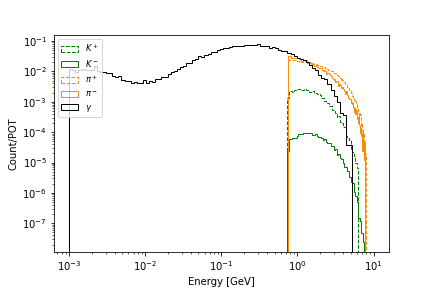}
    \captionsetup{justification=Justified, singlelinecheck=false}
    \caption{Energy flux of photons, positrons, charged pions and kaons produced at BNB}
    \label{fig:fluxbnb}
\end{figure}

\begin{figure}[h]
    \centering
    \includegraphics[width = 0.49\textwidth]{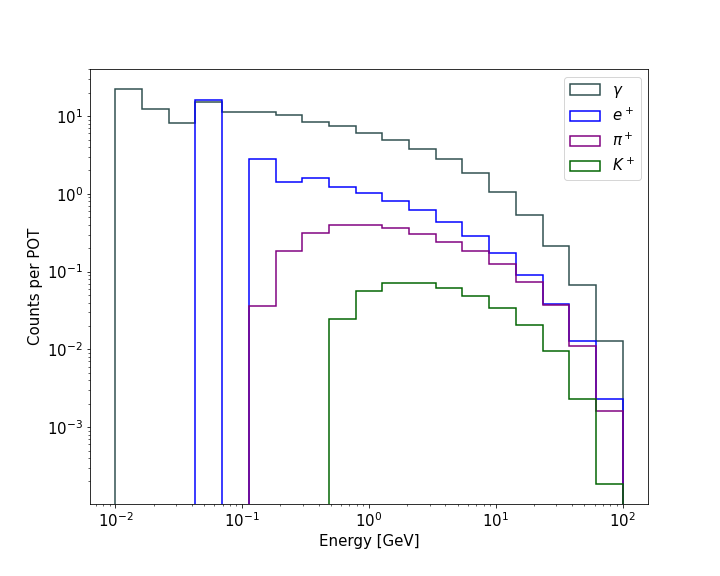}
    \captionsetup{justification=Justified, singlelinecheck=false}
    \caption{Energy flux of photons, positrons, charged pions, and kaons produced at NuMI}
    \label{fig:fluxnumi}
\end{figure}


Figures~\ref{fig:fluxbnb} and \ref{fig:fluxnumi} represent the energy distribution of the secondary particles produced when the 8~GeV and 120~GeV proton beam impinges on the target respectively. It must be noted that the positively charged pions and kaons are along the beamline due to the magnetic horn effect. As for the positrons and photons, they have an angular spread, but the most energetic ones are forward in nature. 

We employ \texttt{GEANT4} to obtain the charged meson fluxes for each of the targets without the effect of the magnetic field. 
To simulate the magnetic horn effect on charged mesons, we apply energy and angle cuts for the charged mesons to match the resulting neutrino fluxes with the reported ones. 
At the BNB target from the neutrino flux, we follow the prescription that only those charged mesons with a total energy greater than 750~MeV, and an angle between [0.03, 0.2] radians to the beamline, pass through the magnet~\cite{Schmitz:2008zz}. 
For LBNF, we follow the empirical models developed in Ref.~\cite{Dev:2023rqb}. We first collect only those particles with kinetic energies greater than 100~MeV and with polar angles between [0.01,1] radians ($z$-axis is the beamline), assigning an additional weight of 3 for every particle with energy less than 10~GeV and 0.2 for those greater than 10~GeV. Using this prescription, we roughly reproduce the muon-neutrino flux~\cite{DUNE:2020ypp} at the DUNE ND. Similar to the flux reported by the DUNE collaboration, our neutrino flux drops down for neutrino energies greater than 10~GeV. On comparing the fluxes for energies below 10~GeV, we notice that there are overestimates and underestimates up to a factor of 3. However, we find that these mismatches do not appreciably affect our sensitivity reaches, especially the ceiling.

\section{Three body decays}

Let us consider a muonphilic scalar $\phi_{\mu}$ emanating from the muon leg in the $K^+$ decay $K^+ (p_0) \rightarrow \nu_{\mu} (p_1) \mu^+ (p_2) \phi_{\mu} (p_3)$, as depicted in Fig.~\ref{fig:leptonscalar}. The coupling strength of the scalar to the muon is $y_{22}$. The matrix element for this process can be expressed as 
\begin{equation}
    \begin{aligned}
        \mathcal{M}_{l,\phi} = -i \frac{G_{F}}{\sqrt{2}}(i y_{22})(\sqrt{2} i {p_0}_{\mu} &V_{us} f_K) \bar{u}(p1)\gamma^{\mu}(1 - \gamma^5)\\
        &\frac{i(\fy{p}_2 + \fy{p}_3) + m_{\mu}}{(p_2 + p_3)^2 - m_{\mu}^2} v(p_2)
    \end{aligned}
    \label{eq:mphilicmel}
\end{equation}

Working with Dalitz variables $m_{12}^2$, $m_{23}^2$, $m_{13}^2$ where, for $i \neq j \neq k$,
\begin{equation}
    m_{ij}^2 = (p_i + p_j)^2 = (p_0 - p_k)^2
\end{equation}
From energy-momentum conservation,
\begin{equation}
    m_{12}^2 + m_{23}^2 + m_{13}^2 = m_K^2 + m_{\nu_{\mu}}^2 + m_{\mu}^2 + m_{\phi}^2
    \label{eq:dalitzfixing}
\end{equation}
The matrix element squared can be written as,
\begin{equation}
    \begin{aligned}
        |\mathcal{M}_{l,\phi}|^2 &= y_{22}^2 \frac{4G_F^2 V_{us}^2 f_K^2}{(m_{23}^2 - m_{\mu}^2)^2}(m_{23}^2 (m_{12}^2 + m_{23}^2 - m_{\phi}^2) \\&
        (m_{23}^2 - m_{\mu}^2) +   
        2 m_{23}^2 m_{\mu}^2 (-m_{23}^2 + m_{K}^2) - \\&(m_{23}^2 - m_{\phi}^2 + m_{\mu}^2) (m_{23}^4 - m_{\mu}^2 m_{\pi}^2))
    \end{aligned}
    \label{eq:mphilicmsquare}
\end{equation}

Although the kaon has the four-momentum given by $(E_K, \Vec{p_K})$, we will work in the rest frame of the decaying kaon after which we can boost the momenta to the lab frame. For a three-body decay, there are 9 ($3 \times 3$) variables determining the momentum (and hence the energy) of the final states. The four constraints on the energy momentum of the final states reduce the degrees of freedom to 5. Out of these 5, three of them determine the plane on which the decay occurs. For the given scalar interactions, the matrix element does not depend on the choice of planes. Hence, they can be integrated, which includes a factor of $2(2\pi)^2$. Therefore, the remaining two degrees of freedom can be chosen to be two of the three Dalitz variables. This will fix the third variable from Eq.~\ref{eq:dalitzfixing}. The differential decay width in the rest frame of the decaying kaon can be written as
\begin{equation}
    \frac{d^2 \Gamma_{0,(l,\phi)}(m_{12}^2, m_{23}^2)}{dm_{12}^2 dm_{23}^2} = \frac{1}{(2\pi)^3 32 m_K^3} |\mathcal{M}_{l,\phi}|^2
    \label{eq:diffdecay}
\end{equation}
Where $|\mathcal{M}_{l,\phi}|^2$ is the matrix element of the process we consider, expressed in Eq.~\ref{eq:mphilicmsquare}. The limits of these integrals are 
\begin{equation}
    \begin{aligned}
        (m_1 + m_2)^2 &\leq m_{12}^2 \leq (m_0 - m_3)^2 \\
        m_{\mu}^2 &\leq m_{12}^2 \leq (m_K - m_{\phi})^2
    \end{aligned}
    \label{eq:m12lims}
\end{equation}
The value of $m_{12}^2$ fixes the energy of particle 3, which is the scalar, in the rest frame of the kaon.
\begin{equation}
    m_{12}^2 = (p_K -p_{\phi})^2 = m_K^2 + m_{\phi}^2 - 2m_K E_{\phi}
    \label{eq:energyphi}
\end{equation}

To integrate over $m_{23}^2$ while fixing the value of $m_{12}^2$, we choose a frame where $p_1 + p_2$ is at rest. Here, the energy, and hence the momentum, of particles 2 and 3 is,
\begin{equation}
    \begin{aligned}
        E_2^* &= \frac{m_{12}^2 + m_2^2 - m_1^2}{2\sqrt{m_{12}}} \\
        E_3^* &= \frac{m_{0}^2 - m_{12}^2 - m_3^2}{2\sqrt{m_{12}}} \\
        p_{2,3}^* &= \sqrt{E_{2,3}^{*2} - m_{2,3}^2}
    \end{aligned}
    \label{eq:m23energies}
\end{equation}

Hence
\begin{equation}
    \begin{aligned}
        (E_2^* + E_3^*)^2 - (p_2^* + p_3^*)^2 &\leq m_{23}^2 \leq (E_2^* + E_3^*)^2 - (p_2^* - p_3^*)^2 \\
        (E_{\mu}^* + E_{\phi}^*)^2 - (p_{\mu}^* + p_{\phi}^*)^2 &\leq m_{23}^2 \leq (E_{\mu}^* + E_{\phi}^*)^2 - (p_{\mu}^* - p_{\phi}^*)^2
    \end{aligned}
    \label{eq:m23lims}
\end{equation}

To generate events, we must randomly choose $m_{12}^2$ between the limits given in Eq.~\ref{eq:m12lims}. This will determine the energy of the scalar in the rest frame of the decaying kaon as per Eq.~\ref{eq:energyphi}. Suppose we choose $n$ values of $m_{12}^2$, then the associated weight for each of these values, and hence the energy, would be
\begin{equation}
    W(m_{12}^2) = \frac{(m_{12,u}^2 - m_{12,l}^2)}{n}\int_{m_{23, l}^2}^{m_{23, u}^2} \frac{d^2 \Gamma_{0,l,\phi}(m_{12}^2, m_{23}^2)}{dm_{12}^2 dm_{23}^2}
\end{equation}

Where the subscripts $u,l$ denote the upper and lower limits of the Dalitz variable, given in Eqs.~\ref{eq:m12lims} and ~\ref{eq:m23lims}. The integral over $m_{23}^2$ is implemented by the trapezoidal rule for integration.

We now obtain a list of $E_{\phi}$, $p_{\phi} = \sqrt{E_{\phi}^2 - m_{\phi}^2}$, and the associated weights in the kaon's rest frame. Since they are isotropic in this frame, the values of $\cos{\theta}$ and $\phi$ can be chosen uniformly between $-1, 1$ and $0, 2\pi$ respectively. By using these and after constructing the four-momentum vector, they can be boosted to the lab frame, which is moving at $(E_K, -\Vec{p_K})$ w.r.t the kaon.

The matrix elements for an HPS emanating from the muon leg are the same as that in Eq.~\ref{eq:mphilicmel} with $y_{22} = \theta m_{\mu}/v$. An additional term where the HPS emanates from the intermediate W boson leg is also involved in this case, which is more dominant than Eq.~\ref{eq:mphilicmel}

\begin{equation}
    \mathcal{M}_{W,\phi} = \frac{G_F}{\sqrt{2}} (\sqrt{2} i {p_0}_{\mu} V_{us} f_K) (2\theta/v) \bar{u}(p1)\gamma^{\mu}(1 - \gamma^5) v(p_2)
    \label{eq:wmel}
\end{equation}

\begin{equation}
    \begin{aligned}
        |\mathcal{M}_{W,\phi}|^2 &= \frac{8G_F^2 V_{us}^2 f_K^2}{v^2}(-m_{12}^2 m_{23}^2 + m_{\mu}^2m_{K}^2 \\
        &+ (-m_{23}^2 +m_{\phi}^2)(m_{23}^2 - m_K^2)) 
    \end{aligned}
    \label{eq:wmelsq}
\end{equation}

Using Eq.~\ref{eq:wmelsq}, the differential branching ratio and the weights can be constructed such as that in Eq.~\ref{eq:diffdecay}.

The three-body decay of a kaon into a gauge boson, such as that in the $U(1)_{T3R}$ model can be constructed similarly. The matrix element for this process is given as
\begin{equation}
    \begin{aligned}
        \mathcal{M}_{l,A'} = -i &\frac{G_{F}}{\sqrt{2}}(i g_{T3R})(\sqrt{2} i {p_0}_{\mu} V_{us} f_K) \bar{u}(p1)\gamma^{\mu}(1 - \gamma^5)\\
        &\frac{i(\fy{p}_2 + \fy{p}_3) + m_{\mu}}{(p_2 + p_3)^2 - m_{\mu}^2}\big( i\gamma^{\nu} \big( \frac{1 + \gamma^5}{2} \big) \big) v(p_2) \epsilon^*_{\nu} (p_3)
    \end{aligned}
    \label{eq:t3rmel}
\end{equation}
Hence, the squared matrix element is 
\begin{equation}
    \begin{aligned}
        |\mathcal{M}_{l,A'}|^2 &= \frac{4G_F^2 V_{us}^2 f_K^2 m_{\mu}^2}{(m_{A'}^2 (m_{23}^2 - m_{\mu}^2)^2)} (m_{A'}^2 (m_{23}^2(m_{\mu}^2 \\
        & - 2(m_{12}^2 + m_{23}^2)) 
        + m_K^2(2m_{23}^2 + m_{\mu}^2)) + (m_{23}^2 \\
        &- m_{\mu}^2) (m_{12}^2 m_{23}^2 - m_{\mu}^2m_K^2) + 2 m_{A'}^4 (m_{23}^2 - m_K^2))
    \end{aligned}
\end{equation}

This matrix element can be used in Eq.~\ref{eq:diffdecay} to generate the $T_{3R}$ gauge boson events.

\section{Bethe-Heitler scattering}

Consider a process where a scalar $\phi$ interacts with nuclei present in the target to produce two leptons, $l^+, l^-$ through the splitting Bethe-Heitler process, as shown in Fig.~\ref{fig:Splitting}. For this process depicted as $N(p_a) \phi(p_b) \rightarrow N(p_1) l^+(p_2) l^-(p_3)$, there are two diagrams that can contribute to it. The matrix element can be written as

\begin{equation}
    \begin{aligned}
        \mathcal{M}_{\text{BH}} &= \bar{u}(p_3)[ \frac{\fy{p_3} - \fy{p_b} + m_l}{(p_3 - p_b)^2 - m_l^2} \gamma^{\mu} + \gamma^{\mu} \frac{\fy{p_b} - \fy{p_2} + m_l}{(p_b - p_2)^2 - m_l^2} ] \\
        & v(p_2) \bar{u}(p_1)\frac{1}{(p_a - p_1)^2}\gamma_{\mu}u(p_a)
    \end{aligned}
\end{equation}

The resulting value of $|\mathcal{M}_{\text{BH}}|^2$ can be calculated using Feyncalc~\cite{Shtabovenko:2016sxi, Shtabovenko:2020gxv}.
The kinematics for a 2 to 3-body scattering involves many more invariants and variables in the differential cross-section. We define five independent dalitz variables in this case. 
\begin{equation}
    \begin{aligned}
        s &= s_{ab} = (p_a + p_b)^2 \\
        s_1 &= s_{12} = (p_1 + p_2)^2 \\
        s_2 &= s_{23} = (p_2 + p_3)^2 \\
        t_1 &= t_{a1} = (p_a - p_1)^2 \\
        t_2 &= t_{b3} = (p_b - p_3)^2 \\
    \end{aligned}
    \label{eq:2to3invariants}
\end{equation}

The other five invariants, $t_{a2}, t_{b2}, t_{a3}, t_{b1}$ and $s_{13}$ can be expressed in terms of those in Eq.~\ref{eq:2to3invariants}.
In order to carry this out, we define the two kinematic functions below.
\begin{equation}
    \lambda (x,y,z) = (x-y-z)^2 - 4yz
    \label{eq:lambdafunc}
\end{equation}

\begin{equation}
    \begin{aligned}
        G(x,y,z,u,v,w) = \begin{vmatrix}
            0 & 1 & 1 & 1 & 1 \\ 
            1 & 0 & v & x & z \\ 
            1 & v & 0 & u & y \\ 
            1 & x & u & 0 & w \\
            1 & z & y & w & 0 \\
        \end{vmatrix}
    \end{aligned}  
    \label{eq:Gfunc}
\end{equation}

The flux factor is given as
\begin{equation}
    F(s, m_a, m_b) = 2(2\pi)^{5} \sqrt{\lambda(s,m_a^2, m_b^2)}
    \label{eq:flux}
\end{equation}

The phase space integral can be expressed as 
\begin{equation}
    \frac{d^5 \Phi}{d\phi dt_1 ds_2 d\Omega_3^{R23}} = \frac{1}{4\sqrt{\lambda(s,m_a^2, m_b^2)}}\frac{\sqrt{\lambda(s_{23},m_2^2, m_3^2)}}{8s_{23}}
    \label{eq:phase}
\end{equation}

Therefore, the differential cross-section can be expressed as 
\begin{equation}
    \frac{d^5 \sigma}{d\phi dt_1 ds_2 d\Omega_3^{R23}} = \frac{1}{F(s, m_a, m_b)}\frac{d^5 \Phi}{d\phi dt_1 ds_2 d\Omega_3^{R23}} |\mathcal{M}_{\text{BH}}|^2 
    \label{eq:diffcs}
\end{equation}
Where $d\Omega_3^{R23} = d\cos{\theta_{b3}^{R23}}d\phi_3^{R23}$, which are the angles in the rest frame of particles 2 and 3, which is $l^+,l^-$. These angles can be expressed in terms of the Dalitz invariants as below.
\begin{equation}
    t_2 = t_{b3} = m_b^2 + m_3^2 - 2E_b^{23}E_3^{23} + 2p_b^{23}p_3^{23}\cos{\theta_{b3}^{23}}
    \label{eq:t2}
\end{equation}

\begin{equation}
    \begin{aligned}
        s_{12} &= s + m_3^2 - 
        \frac{1}{\lambda(s_{23}, t_{a1}, m_b^2} 
        \bigg( 
        2\sqrt{G(s, t_{a1}, s_{23}, m_a^2, m_b^2, m_1^2)} \\
        &\sqrt{ G(s_{23}, t_{b3}, m_3^2, t_{a1}, m_b^2, m_2^2)}\cos{\phi_b^{23}} + \\    
        &\begin{vmatrix}
                2m_b^2 & s_{23} - t_{a1} + m_b^2 & m_b^2 + m_3^2 - t_{b3} \\ 
                s_{23} - t_{a1} + m_b^2 & 2s_{23} & s_{23} - m_2^2 + m_3^2 \\ 
                s-m_a^2 + m_b^2 & s + s_{23} - m_1^2 & 0 \\ 
            \end{vmatrix}
        \bigg)
    \end{aligned}
\end{equation}

The limits of the integral are
\begin{equation}
    \begin{aligned}
        (m_2 + m_3)^2 &\leq s_{23} \leq (\sqrt{s} - m_1)^2 \\
        -1 &\leq \cos{\theta_{b3}^{R23}} \leq 1 \\
        0 &\leq \phi_3^{R23} \leq 2\pi \\
        m_a^2 + m_1^2 - &2E_{a}^{23}E_{1}^{23} - 2p_{a}^{23}p_{1}^{23} \leq t_{a1} \\
        &\leq m_a^2 + m_1^2 - 2E_{a}^{23}E_{1}^{23} + 2p_{a}^{23}p_{1}^{23} 
    \end{aligned}
\end{equation}

The energies in the rest frame of $23$ are as given below,
\begin{equation}
    \begin{aligned}
        E_a^{23} &= \frac{s_{ab} + m_a^2 - m_{b}^2}{2\sqrt{s_{ab}}} \\
        E_1^{23} &= \frac{s_{ab} + m_1^2 - s_{23}}{2\sqrt{s_{ab}}} \\
        E_b^{23} &= \frac{s_{23} + m_b^2 - t_{a1}}{2\sqrt{s_{23}}} \\
        E_3^{23} &= \frac{s_{23} + m_3^2 - m_{2}^2}{2\sqrt{s_{23}}} \\
    \end{aligned}
\end{equation}
The momenta can be obtained from the above energies. By integrating these variables, we obtain the total cross-section of the 2-3 Bethe Heitler scattering process for a given energy and mass of the incoming scalar. The total number of Bethe-Heitler scattering events is given by

\begin{equation}
    N_{\text{BH}}(m_{\phi}) = \int dE_{\phi} \frac{dN_{\phi}}{dE_{\phi}}\frac{N_{A}\rho_{\text{T}}}{A_{\text{T}}}L_{\text{det}} \sigma_{\text{BH}}(dE_{\phi}, m_{\phi})
\end{equation}
Where $N_A$, $\rho_{\text{T}}$, and $A_{\text{T}}$ are the Avogadro number, density, and atomic mass of the detector respectively. $L_{\text{det}}$ is the length of the detector.

\newpage
\bibliography{references}

\end{document}